\documentclass[format=acmsmall, review=false, screen=true,table]{acmart}

\AtBeginDocument{%
  \providecommand\BibTeX{{%
    \normalfont B\kern-0.5em{\scshape i\kern-0.25em b}\kern-0.8em\TeX}}}

\setcopyright{acmcopyright}
\acmJournal{PACMHCI}
\acmYear{2021} \acmVolume{5} \acmNumber{CSCW2} \acmArticle{403} \acmMonth{10} \acmPrice{15.00}\acmDOI{10.1145/3479547}

\usepackage{booktabs} 
\usepackage[ruled]{algorithm2e} 

\SetAlFnt{\small}
\SetAlCapFnt{\small}
\SetAlCapNameFnt{\small}
\SetAlCapHSkip{0pt}
\IncMargin{-\parindent}

\usepackage{url}      
\usepackage{color, colortbl}
\usepackage{tabularx}
\usepackage{textcomp}
\usepackage{booktabs}
\usepackage{todonotes}
\usepackage{amsmath}
\usepackage{graphicx}
\usepackage{subfigure}
\usepackage{subfigure}
\usepackage{comment}
\usepackage{pbox}
\usepackage{tabu}
\usepackage{hyphenat}
\usepackage{makecell}
\usepackage{multirow}
\usepackage{fancyhdr}

\newcommand{\revise}[1]{\textcolor{black}{#1}}
\newcommand{\newrevise}[1]{\textcolor{black}{#1}}
\newcommand{\chen}[1]{\textcolor{blue}{CHEN: #1}}
\newcommand{\suyu}[1]{\textcolor{yellow}{Suyu: #1}}

\newcommand{\tool}{\texttt{MathLatexEdit}} 
\newcommand{\site}{Mathematics Stack Exchange }

\begin{document}

\title{Latexify Math: Mathematical Formula Markup Revision to Assist Collaborative Editing in Math Q\&A Sites}


\author{Suyu Ma}
\affiliation{%
  \institution{Monash University}
  \department{Faculty of Information Technology}
  \city{Melbourne}
  \country{Australia}
}

\email{suyu.ma1@monash.edu}

\author{Chunyang Chen}
\authornote{Corresponding author.}
\affiliation{%
  \institution{Monash University}
    \department{Faculty of Information Technology}
  \city{Melbourne}
  \country{Australia}
}
\email{chunyang.chen@monash.edu}

\author{Hourieh Khalajzadeh}
\affiliation{%
  \institution{Monash University}
    \department{Faculty of Information Technology}
  \city{Melbourne}
  \country{Australia}
}
\email{hourieh.khalajzadeh@monash.edu}

\author{John Grundy}
\affiliation{%
  \institution{Monash University}
    \department{Faculty of Information Technology}
  \city{Melbourne}
  \country{Australia}
}
\email{john.grundy@monash.edu}

\renewcommand{\shortauthors}{Suyu Ma et al.}

	\begin{abstract}
	Collaborative editing questions and answers plays an important role in quality control of \site which is a math Q\&A Site. 
	Our study of post edits in \site shows that there is a large number of math-related edits about latexifying formulas, revising LaTeX and converting the blurred math formula screenshots to LaTeX sequence.
	Despite its importance, manually editing one math-related post especially those with complex mathematical formulas is time-consuming and error-prone even for experienced users.
	To assist post owners and editors to do this editing, we have developed an edit-assistance tool, {\tool} 	for formula latexification, LaTeX revision and screenshot transcription.
	We formulate this formula editing task as a translation problem, in which an original post is translated to a revised post.
	{\tool} implements a deep learning based approach including two encoder-decoder models for textual and visual LaTeX edit recommendation with math-specific inference.
	The two models are trained on large-scale historical original-edited post pairs and synthesized screenshot-formula pairs.
	Our evaluation of {\tool}  not only demonstrates the accuracy of our model, but also the usefulness of {\tool} in editing real-world posts which are accepted in Mathematics Stack Exchange.

	\end{abstract}

\ccsdesc[500]{Applied computing~Text editing}
\ccsdesc[300]{Human-centered computing~Collaborative and social computing}

%
%

\keywords{Q\&A sites, Collaborative editing, Deep learning, Latex, Math}

\maketitle

\section{Introduction}
\label{intro}
As the foundation of science and engineering, mathematics has always been one of the most important subjects for students~\cite{JenlinkKarenEmbry2006ME} among all levels of education.
It can help reshape the students' reasoning, creativity, critical thinking and problem-solving abilities.
On the other hand, mathematics is also one of the most challenging subjects for students ~\cite{JenlinkKarenEmbry2006ME, MiddendorfJessica2018IRtM, DanesiMarcel.author2016LaTM}.
Various mathematics community Question and Answer (Q\&A) sites have been developed.
They provide a platform for students to ask about anything related to the mathematics with fast feedback from peers and experts.
Mathematics Stack Exchange~\cite{math} is an example of such a Q\&A site for people studying mathematics at any level and for professionals in related fields.
Since its launch in 2010, it contains 1.3 million questions and 1.7 million answers from 661 thousand users~\cite{math2}. 

However, the dramatic growth of posts and users on such Q\&A sites poses a severe challenge to the quality assurance for site content.
For example, some novice users may post questions and answers with grammar errors, misuse of abbreviations, blurred screenshots of formulas, and the lack of important information (e.g., context or referenced resources) for understanding questions and answers.
Such quality deterioration can negatively influence the readability and understandability of posts, and may further discourage the participation of users~\cite{li2015good}.
To avoid the quality decay of the sites, Q\&A sites like \site provide official recommendations for effective question writing~\cite{help3} and answer writing~\cite{help1}.

Although all users are encouraged to follow the quality assurance policies, many users may still violate them carelessly or unintentionally, and some users may not even be aware of the existence of such policies. 
To ensure the site quality~\cite{MamykinaLena2011Dlft}, \site encourages users, especially experienced users~\cite{help2}, to collaboratively edit the posts to make them comply with the site quality standards.
According to our analysis in Section~\ref{sec:empiricalStudy}, among 2,886,174 posts including questions and answers (as of Mar 2020), 1,318,753 (45.69\%) of them have been edited at least once.
Post edits involve not only minor corrections of misspellings and
grammar errors, but also math-related such as formatting math formula for better readability, fixing mathematical mistakes.

Although collaborative editing is beneficial for the community~\cite{Li2016PredictingCE}, there are still three problems with such mechanism. 
First, it requires significant community effort, especially from high-reputation users to edit the posts directly and/or approve the edits by the other users. 
Second, some errors in the original posts, especially relatively complicated ones such as math formula errors or format are difficult to spot, as they may require a good understanding of
the question or answer content. 
Third, all these collaborative edits are reactive to existing errors which may have already harmed the readers of the posts before edits, or made it difficult for those who want to help to answer the questions.

Therefore, in addition to collaborative editing for reactive quality assurance, we also need a more proactive mechanism of quality assurance which could check a post before it is posted, spot the potential issues in the post, and remind the post owner to fix issues if any.
Towards that target, there are some research works investigating the automated revision of Q\&A posts~\cite{chen2017community, chen2018data, Li2016PredictingCE, vargo2016editing}.
However, different from other Q\&A sites (e.g., Quora\footnote{\url{https://www.quora.com/}} and Stack Overflow\footnote{\url{https://stackoverflow.com/}} ), there are some domain-specific edits required on \site due to the mathematics characteristics which cannot be addressed by their automated methods.
By analysing the topics in the comments of post edits and empirical study of historical math-related edits in Section~\ref{sec:empiricalStudy}, three editing types emerge including formula latexification, LaTeX formula revision, formula screenshot transcription, involving 516,010 related edits.
Considering the wide range of post content and formats involved in post edits (see Table~\ref{tab:access} for examples), it would require significant manual effort to develop a complete set of rules for representing editing patterns.

This challenge motivates us to develop a data-driven deep-learning based janitor for recommending post owners or other users potential math-related edits, by automatically learning from historical edits.
Note that \site encourages users to use LaTeX, a computer programming language used in typesetting technical documents~\cite{math_latex}, for writing mathematical formulas.
It can be rendered to clear and nice-looking formulas by Mathjax~\cite{CervoneDavide2012MAPf}.
In this work, we formulate the post editing recommendation to a translation task, in which an original post is ``translated'' into the edited post.
This includes translating erroneous/unformatted math formula to the correct/formatted one (textual LaTeX edit) and math formula screenshots to more readable ones in LaTex (visual Latex edit). 
\revise{
We design two models for these two tasks and define them as textual edit model and visual edit model.
We adopt transformer-based model for textual edits and use DenseNet~\cite{huang2017densely} and Long Short-term Memory (LSTM)~\cite{hochreiter1997long-lstm} models for visual edits.
To improve the performance of our tool, we adopt the sentence normalization which shortens the length of the input by directly copying the non-math content when fixing formula error and formula formatting (Section~\ref{textual_edit}).}
We also improve the basic screenshot transcription method by adding image similarity into the inference phase for boosting the quality of generated LaTex.
\revise{We name our tool as Mathematical Latex Editor (MathLatexEdit)}.

The trained {\tool} automates the math-specific edits, thus removing the need for prior manual editing rule development.
\revise{Our work is the first one to explore how to support collaborative editing in math-related question and answer sites.  
Our work fills in the gap between math-related post editing and cooperative work.
This tool will not only help post owners reduce minor math-related issues before posting their questions and answers, but also help post editors improve their editing efficiency. 
Furthermore, the identified issues together with the recommended corrections will help novice post editors learn community-adopted editing patterns.}


To train and test our model, we develop a text differencing algorithm to collect a large dataset of original-edited sentence pairs from post edits in \site for textual edit model and prepare a large dataset of LaTeX formula and image pairs for visual edit model.
Our results show that our approach outperforms other rule-based or deep learning based baselines for post edit recommendation.
Apart from the model performance, we also conducted \revise{a user study to confirm that participants with our tool can finish the editing with less time but higher precision, recall and satisfactoriness.}
The field study also shows its usefulness in which the first author acts as a novice post editor that has little post editing experience. 
Based on the post editing recommendations by our tool, he edited 80 posts and submit the edits to Mathematics Stack Exchange, and 78 of them were accepted. 
That is, for each accepted post edit, at least two trusted contributors considered that the edit had significantly improved the post quality.

We make the following key contributions in this paper:
\begin{itemize} 
\item We conducted an empirical study of collaborative post editing in \site including the editing types and editing content, identifying the need for an edit assistance tool for mathematics formulae.

\item We developed {\tool}, a deep learning based edit assistance tool that can latexify math expression, revise LaTeX formulas, convert formula screenshots in \site to LaTeX sequence for further MathJax rendering to improve the post quality and readability. \revise{Although the current work studies {\site}, our data analysis method and deep learning approach could be applied to other social systems which is discussed in Section~\ref{sec:generalization}}.

\item We evaluated {\tool} from two perspectives, including the quality of post edit recommendation based on a large-scale dataset and the usefulness to assist a novice post editor in editing unfamiliar new posts. 
\end{itemize}

\section{Related Work}


\subsection{Mathematics Q\&A Sites}
Online question and answer (Q\&A) sites are platforms for participants to ask and answer questions.
With the power of crowd-sourced answers, Q\&A sites, such as Quora\footnote{\url{https://www.quora.com/}} and Stack Overflow\footnote{\url{https://stackoverflow.com/}}, are more and more popular with accumulating millions of questions and answers in different domains~\cite{ma2019easy,chen2017unsupervised,chen2019sethesaurus,cao2021automated}. 
Due to the importance and challenges of learning math, math-specific Q\&A sites have also been launched, such as \site\footnote{\url{https://math.stackexchange.com/}} for students and MathOverflow\footnote{\url{https://mathoverflow.net/}} for professional and expert researchers.
MathOverflow has attracted great attention from researchers.
Montoya et al. \cite{montoya2013social} model MathOverflow as a social network for analyzing its social achievement and centrality.
Tausczik et al. \cite{tausczik2014collaborative} investigate the collaboration patterns when solving a research-level mathematical problem in MathOverflow. 

However, few researchers explore the Mathematics Stack Exchange. 
As one of the most popular mathematics Q\&A sites with 635 new questions and 200K visits per day~\cite{math2}, \site deserves more attention from our research community.
To enhance the post quality in Mathematics Stack Exchange, our study is the first work to explore its collaborative editing patterns and propose a deep learning method to assist post owners and editors to revise low-quality posts.

\subsection{Collaborative Editing in Q\&A Sites}
Although collaborative editing is widely studied in some user-generated content (UGC) communities (e.g., Wikipedia~\cite{Wiki},  wikiHow~\cite{wiki:how}), there are relatively few works focusing on Q\&A sites.
Collaborative editing can help improve the quality control in Q\&A sites by converting  low-quality posts to higher-quality ones~\cite{MamykinaLena2011Dlft}.
Li et al.~\cite{li2015good} demonstrate that it will not hurt the user engagement, despite the quality improvement.
Ford et al.~\cite{ford2018we} show that collaborative editing with mentors can further improve engagement in Q\&A site.
Choi and Yla~\cite{choi2018will} find that the moderators in collaborative editing can help resolve the conflict of using tags to describe the questions which is caused by users' background and understanding differences.

Although collaborative editing is important for the site quality, it takes much human effort to realise.
To assist collaborative editing in online communities, some machine learning based methods have been proposed.
Li et al.~\cite{Li2016PredictingCE} and Chen et al.~\cite{chen2018data} developed a classifier to automatically predict if a post needs edits or what type of edits are required e.g. adding links, images, updating format, etc.
Chen et al.~\cite{chen2017community} proposed a deep learning based model to automate some minor revisions such as misspelling, grammar errors and keyword formatting in sentences of the post.

Our {\tool} tool differs in two key aspects.
First, most related works above focus on the most popular programming Q\&A site, Stack Overflow~\cite{MamykinaLena2011Dlft, li2015good,vargo2016editing,chen2017community}, while we focus on the Mathematics Stack Exchange.
We carried out an empirical study to explore the domain-specific edits in this math-related Q\&A site.
Second, in addition to their models based on the text of posts, we are also working on a more challenging type of collaborative editing, i.e., converting formula screenshots to LaTeX sequence that bridges the gap between visual and textual information.

\subsection{Grammar Error Correction}
We formulate the post editing recommendation in this work as a translation task i.e., translating the text, buggy latex commands and formula screenshots to corresponding correct latex commands.
There are many research works on automated textual grammar error correction with machine learning~\cite{junczys2016phrase, mizumoto2016discriminative, yuan2016candidate} i.e., detecting and fixing grammar error of the original natural-language text.
For example, Junczys-Dowmunt and Grundkiewicz design an approach that can automatically correct grammar errors with phrase-based Statistical Machine Translation (SMT) method~\cite{junczys2016phrase}.
Mizumoto and Matsumoto \cite{mizumoto2016discriminative} and Yuan et al. \cite{yuan2016candidate} recommend ranked grammar error correction with SMT method and a ranking method.
Unlike traditional SMT methods, Neural Machine Translation (NMT), such as RNN-based methods and transformer-based methods~\cite{vaswani2017attention}, utilize sentence context information and joint all the components in the training process.
CNN based Seq2Seq and quality estimation methods are used by Chollampatt and Ng~\cite{chollampatt-ng-2018-neural} to automatically estimate the quality of GEC sentences.
Grundkiewicz and Junczys-Dowmunt combine SMT and neural machine translation to automated Grammatical Error Correction  ~\cite{grundkiewicz-junczys-dowmunt-2018-near}.
Different from their works, we are the first to target at mathematic-specific revisions i.e., the content change of latex commands by taking the mathematic characteristic into the consideration.

\subsection{Mathematics Formula Accessibility}
It is crucial for the community to make mathematical formula accessible~\cite{MiddendorfJessica2018IRtM, DanesiMarcel.author2016LaTM}, as many users just take a screenshot of mathematics formula and put it online which can significantly reduce the readability of normal users and engagement of users with vision impairment.
To assist the conversion of mathematics formula screenshots to relevant representation in LaTeX, researchers propose many different algorithms. 
A commercial OCR tool is used by Garain et al. ~\cite{garain2004identification} to classify text, and the unrecognized patterns were further analyzed to detect mathematics formulas. 
Based on larger symbols and blank spaces, T waaliyondo el al.~\cite{twaakyondo1995structure} divide the formulas into sub expressions and represent them as a tree.
Suziki et al.~\cite{garain2004identification} use a similar approach with a minimum cost spanning-tree algorithm. 
The commercial software InftyReader~\cite{SuzukiMasakazu2003IaiO-infty} is based on this work. 
Inspired by the image captioning tasks~\cite{xu2015show, chen2018ui, chen2020unblind}, Deng et al.~\cite{deng2016you} designed a more advanced deep learning based model.

\revise{Compared with these general works, our image transcribing algorithm is the first targeting at collaborative editing in the real Mathematics Stack Exchange.}
Our approach not only takes the textual Latex edit into  consideration and is also more advanced by incorporating DenseNet and inference with additional visual similarity for converting screenshot into Latex representation described in Section~\ref{sec:method}.
\revise{Different from these theoretical works only focusing on the model performance on testing dataset, we also carried out a user study and a field study in Section~\ref{sec:usefulness} to verify the usefulness in assisting with real-world post edits.}
\section{COLLABORATIVE EDITING ANALYSIS OF \site}

\label{sec:empiricalStudy}
\revise{Different from other Q\&A sites (e.g., Quora and Stack Overflow), there are some domain-specific edits required on Mathematics Stack Exchange due to the mathematics characteristics which cannot be addressed by automated methods.}
We downloaded the latest data dump\footnote{https://archive.org/download/stackexchange} of \site which contains 2,886,174 posts (including 1,216,368 questions and 1,669,806 answers) and all post edits since its launch on July 20, 2010 to March 1, 2020.
Based on this large dataset, we carried out an empirical study of post edits in \site to understand the characteristics of post editing in \site and to motivate the required tool support.
\newrevise{
This empirical study allows us to re-frame our research from the perspectives of quality control and beneficial aspects of our tool. }



\subsection{What are the edits about?}
In Mathematics Stack Exchange, there are three kinds of post information which can be edited -- question tags~\cite{chen2016mining, chen2016techland}, question title, and post (question and answer) body~\cite{MamykinaLena2011Dlft}.
Question-title and post-body editing are of the same nature (i.e., sentence editing), while question-tags editing is to add and/or remove the set of tags of a question.


As of March 1, 2020, there have been in total 2,696,115 post edits.
Among them, 334,136 (12.39\%) are question-title edits,  286,715 (10.63\%) are question-tag edits, and the majority of post edits (2,075,264 (76.97\%)) are post-body edits.
The tags of 268,620 (22.08\%) questions, the titles of 243,184 (20.00\%) questions, and the body of 1,075,420 (37.26\%) posts have been edited at least once.
63.21\% of these edits are self-edits by the post owners and the other 36.79\% are collaborative edits by other post editors.
Overall, post-body edits make up the majority of post edits, and post-body editing are more complex compared with revising title or question tags.
Therefore, we focus on post-body edits in this work.
Hereafter, post edits refer to post-body edits, unless otherwise stated.

\subsection{
Who edited posts?
}
\revise{
Among all 2,075,264 post  edits, 1,228,763 (59.21\%) are self-edits by the post owners, 846,501 (40.79\%) are edits by other users.
This data suggests that an edit assistance tool may benefit the \site community from three perspectives.
}

\revise{
First, the tool can highlight issues in posts that post owners are creating and proactively remind them to fix the issues. 
This reduces the need for editing after the creation and ensures the post quality in the first place.
Second, an edit assistance tool that recommends edits can improve the efficiency in editing others' posts by providing a reliable starting point.
Third, the post editors can use our tool to learn to correct minor domain-specific issues.
Such small editing tasks can provide a mechanism of legitimate peripheral participation \cite{lave1999legitimate} for novice users who have no experience in {\site}.
This may help to on-board novice users in \site and improve the quality of their post edits.
}

\subsection{What has been edited?}
\label{sec:editTopic}
To understand what post edits are about, we analyzed the comments attached to the post edits. 
In Mathematics Stack Exchange, when users finish editing a post, they can add a brief comment to explain  the editing reasons. 
We collected all post-edit comments and applied standard text processing steps to post-edit comments such as removing punctuations, lower-casing all characters, and excluding stop words.

\begin{table}
	\caption{Topics of post-edit comments}
	\small
	\begin{center}
		\begin{tabular}{l|l|l}
		\toprule

			ID&Topic Name&Keywords\\

			\hline
			1&spelling \& grammar & grammatical error, spelling, typo\\
            2&adding links&wikipedia link, broken link, linked\\
			3&explaining the question&descriptive, explanation, definition\\
			4&converting image&replace image, latexifying image, transcribed image\\
			5&revising mathematical content&algebraic, mathematical, mathematics \\

			6&changing format&adjust format, formatting, formatted\\
			7&improving readability &improve readability, easier read, make readable\\
			8&modifying LaTeX formula&tex, improving latex, tex improvement, applied mathjax \\
		\bottomrule

		\end{tabular}
		\label{tab:LDA}
	\end{center}
	\vspace{-2mm}
\end{table}

We extracted the topics from users' editing comments to understand their editing content.
To extract common editing types, we adopted the Latent Dirichlet Allocation (LDA)~\cite{blei2003latent-lda} model to analyze the post-edit comments. 
LDA is a statistical model for discovering abstract topics that occur in a collection of documents in which each topic consists of a set of keywords. 
A significant limitation of LDA is that it considers only single words (i.e., unigrams). 
However, a single word may not capture the exact semantics of the post-edit comments. 
In contrast, phrases that are composed of several words are more intuitive to understand the intention behind post edits, such as ``latexifying image'' instead of ``image'', ``improving latex'' rather than ``latex''.
Therefore, these multi-word phrases must be recognized and treated as a whole in LDA model. 

We adopted a simple data-driven and memory-efficient approach~\cite{mikolov2013distributed} to detect multi-word phrases in post-edit comments. 
In this approach, phrases are formed iteratively based on the unigram and bigram counts, using the following formula $score(w_i,w_{i+1})=\frac{count(w_i,w_{i+1})-\theta}{count(w_i) \times count(w_{i+1})} \times N$.
The $w_i$ and $w_{i+1}$ are two consecutive words. $\theta$ is a discounting coefficient to prevent infrequent bigrams to be formed. 
That is, the two consecutive words will not form a bigram phrase if they appear as a phrase less than $\theta$ times in the corpus. 
N is the vocabulary size of the corpus. 
In this work, we experimentally set $\theta$ as 10 and the threshold for score as 10 to achieve a good balance between the coverage and accuracy of the detected multi-words phrases.

Our method can find bigram phrases that appear frequently enough in post-edit comments compared with the frequency of each unigram, such as ``improve readability''. 
However, bigram phrases like ``it is'' will not be formed because each unigram also appears very frequently in the text. 
All these phrases are then concatenated with underline like ``improve\_readability'' in the corpus of post-edit comments, and then we use the LDA model to extract the topics.

We extracted 8 topics with corresponding keywords shown in Table \ref{tab:LDA}. 
Note that we annotated the topics name with our own summarization based on the topic keywords. 
Different from individual frequent keywords, these topics provide finer-grained information.
Apart from common types mentioned above, there are also some domain-specific editing patterns related to mathematics formula (\#5 revising mathematical content, \#8 modifying LaTeX formula) and readability (\#4 converting image, \#6 changing format, \#7 improving readability).
These two common editing types represent the community norms that \site commits its effort to maintain.


\subsection{What efforts have been committed to the domain-specific edits? }
\label{sec:editType}
\begin{table}
	\caption{Types of domain-specific accessible edits 
	}
	\begin{center}
		\begin{tabular}{l}
		\toprule
		1 changing formula format\\
			\includegraphics[scale=0.13]{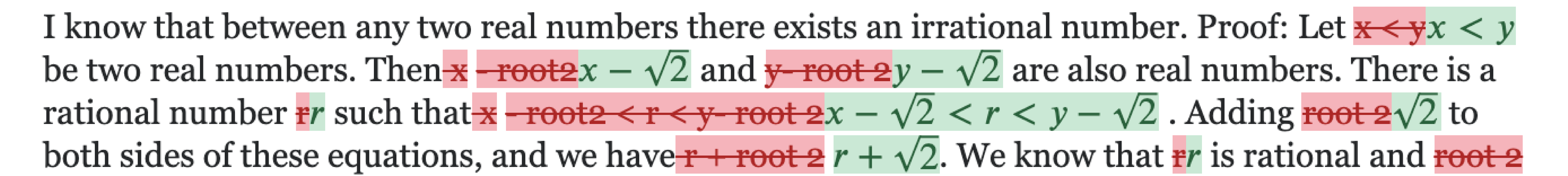}\\
			\hline
			2 correcting mistakes in formula\\
			\includegraphics[scale=0.35]{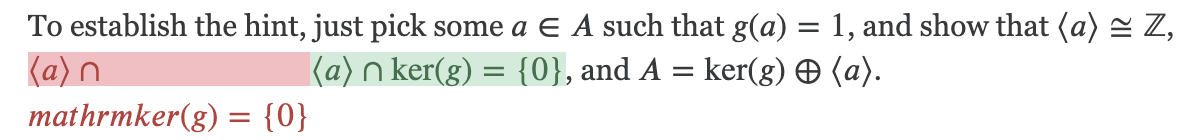}
			\includegraphics[scale=0.35]{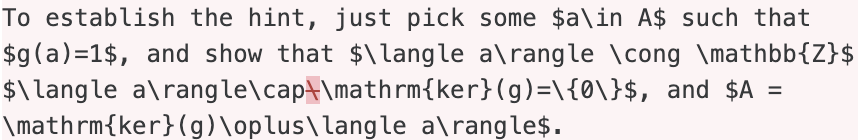}\\
			\hline
			
			3 replacing image with LaTeX formula\\
            \includegraphics[scale=0.3]{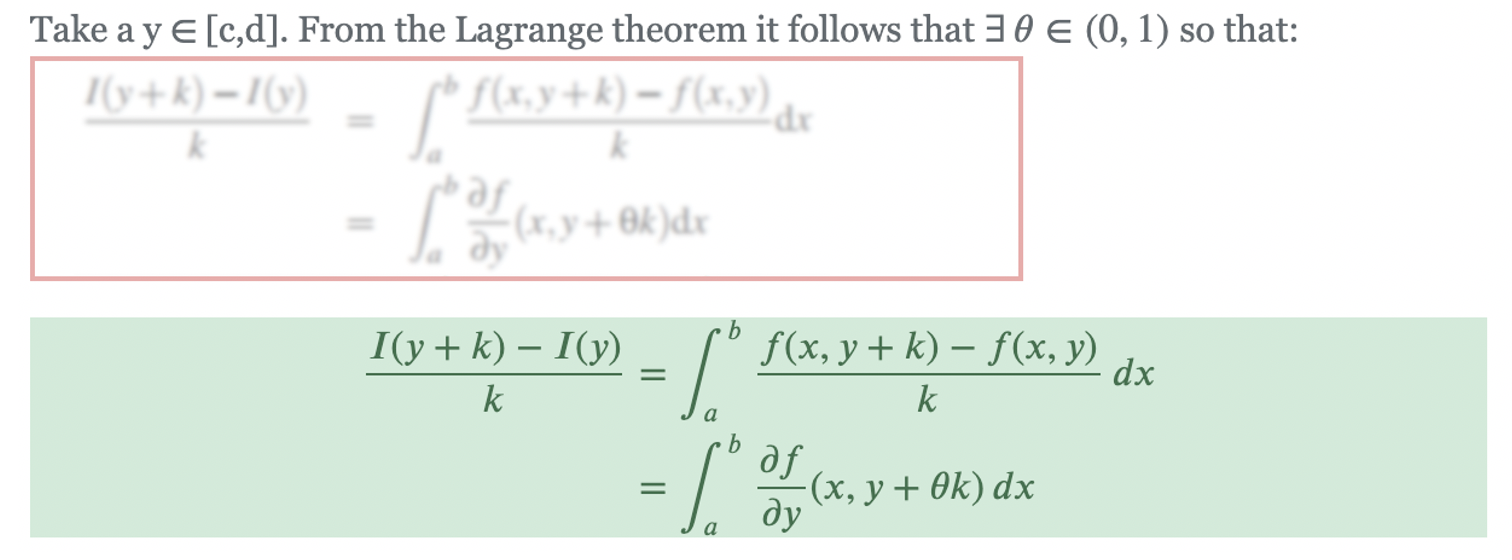}\\
		\bottomrule

		\end{tabular}
		\label{tab:access}
	\end{center}
	\vspace{-2mm}
\end{table}
\revise{
As mentioned in Section~\ref{sec:editTopic}, many post edits are domain-specific and highly related to mathematics, especially mathematics formula and readability, according to the analysis of the editing comments.}
To further explore what edits are about, we extracted the edits with editing comments including domain-specific words such as ``formula'', ``LaTeX'', ``math'', etc.
By traversing the commented editing history in Mathematics Stack Exchange, we collected 155,369\footnote{\newrevise{Note that only 726,342 edits contain comments, so the number is highly underestimated.}} domain-specific edits in total.
We then randomly selected 100 of them for manual inspection.
According to our observation, there are three common mathematics domain-specific editing patterns, as seen in Table~\ref{tab:access}.

First, 29 edits are to change a plain formula into LaTeX format with MathJax rendering for a clearer view, resulting in better readability. 
It may be the first time for some users to join the Mathematics Stack Exchange, so they write a plain math formula in the post.
But editing it into LaTeX can distinguish the mathematics formula from the plain text so that other users can easily understand the meaning of the post, leading to higher possibility of responses.
For instance, ``x - root2'' is converted into ``$x-\sqrt{2}$'' in the first example of Table~\ref{tab:access}.
Compared with natural English words, LaTeX mathematical expression highlights these numerical tokens and help readers easily capture the key point of the posts. 

Second, 35 edits focus on correcting mistakes in formulas in order to clarify the content.
Some tokens in formulas are missed or misspelled. 
Even the experienced users may make mistakes when writing math formulas, especially for the complicated ones.
Since the math formula may be the most important content, many edits are targeting at it for more accurate information.
For instance, 
 extra $\backslash$ is removed in second example of Table~\ref{tab:access}, which improves the quality of the post.

Third, 8 edits are to convert the blurred formula images into LaTeX which is further rendered to a vector graph so that users can zoom in or out for better view in the third example of Table~\ref{tab:access}.
There are mainly two kinds of blurred formula images, which includes screenshots and pictures captured by cameras.
Compared with the blurred formula images, the LaTeX format also makes it easy to be indexed by the search engine, resulting in the searchable text.

The rest edits are not related to mathematics, such as adding or deleting the content in posts, correcting the grammar errors in posts, or changing the display formats.

\subsection{How many mathematic-specific edits are there?}
In Mathematics Stack Exchange, all mathematical formulas should be written in LaTeX format with further MathJax rendering for clear visualization.
\revise{For each of the three types of mathematics-specific edits (formula latexification, LaTeX revision, and screenshot transcription) appear most often in the last section, we observe the corresponding detailed text changes for detecting instances of a particular type of edits and their frequencies as follows}:
\begin{itemize}
	\item To render better readability of mathematic formula, \site encourages users to annotate their formula in LaTeX with $\$$ at the beginning and end (Table~\ref{tab:access}.(1)). 
	Users also need to use the LaTeX grammar like some special symbols. 
	The revision from plain math formula in text into LaTeX is called as \textit{formula latexification}.
	\item To improve the accuracy of formulas, users are encouraged to revise others' formula LaTeX in posts which is called \textit{latex revision} (Table~\ref{tab:access}.(2)) in this work.
	\item To improve the readability and accessibility of math formula, the embedded image/screenshots which require a special link (Table~\ref{tab:access}.(3)) ending with postfix including .png, .jpg, .gif, .bmp and .tiff need to be converted to formula in LaTeX with $\$$ at the beginning and end. We refer to this kind of revision of screenshot links as \textit{screenshot transcription}.
\end{itemize}

\revise{By differencing the original post body and the edited post body and following last three patterns, we count the number of different types of edits. 
There exist 627,251  math-specific edits in total, in which 516,010 (24.86\% out of all post body edits) edits are following three types.
}
\begin{itemize}

	\item 169,006 (169,006 / 516,010 = 32.75\%) post edits include \textit{formula latexification}.
	\item  288,947 (55.99\%) post edits include \textit{LaTeX revision}, and 548,230 formulas are revised in total.
	\item For \textit{screenshot transcription}, 59,141 (11.46\%)  post-body edits include  that conversion.
\end{itemize}


\revise{Considering the diversity of text, formula, pixels and context involved in math-related edits, it would require significant manual effort to develop and validate a complete set of rules for representing editing patterns. For example, ``root'' should be converted to ``\textbackslash sqrt'' in latex, but will not be changed in many other contexts. So, it is impractical to enumerate all such cases.
An advanced approach is highly needed for automated editing.}

\textbf{Summary}: 
45.69\% posts in \site have been edited involving a variety of editing types, including fixing grammatical errors, clarifying the meaning of a post, formatting the post and adding related resources or hyperlinks. 
In addition, there are also many domain-specific edits like formula latexification, LaTeX revision,  screenshot transcription for better readability and accessibility.
\revise{
These 516,010 math-specific edits (24.86\% out of all post body edits) require much more human effort to maintain in the community due to the complexity of math formulas.
Therefore, a post edit recommendation algorithm is needed to assist users in \site to correct math-specific errors and assist these three math-related edits for ensuring the post quality.
}
\section{Recommending Latex Edits by Deep Neural Network}
\label{sec:method}
The three types of post edits related to math formula in our empirical study highlight the community efforts for ensuring the post quality in Mathematics Stack Exchange.
Unfortunately, these efforts and revisions are implicit knowledge in millions of post edits.
Considering the diversity of post editing types and contexts, it would require significant human effort to build a complete set of rules to deal in all different situations.
Therefore, we developed a deep-learning based approach which can automatically improve the post editing patterns from historical post edits, and recommend edits to the new posts based on the learned editing knowledge.

\subsection{Overview of our \textit{{\tool}} Approach}
\label{overview}

\begin{figure}
	\centering
	\includegraphics[scale=0.35]{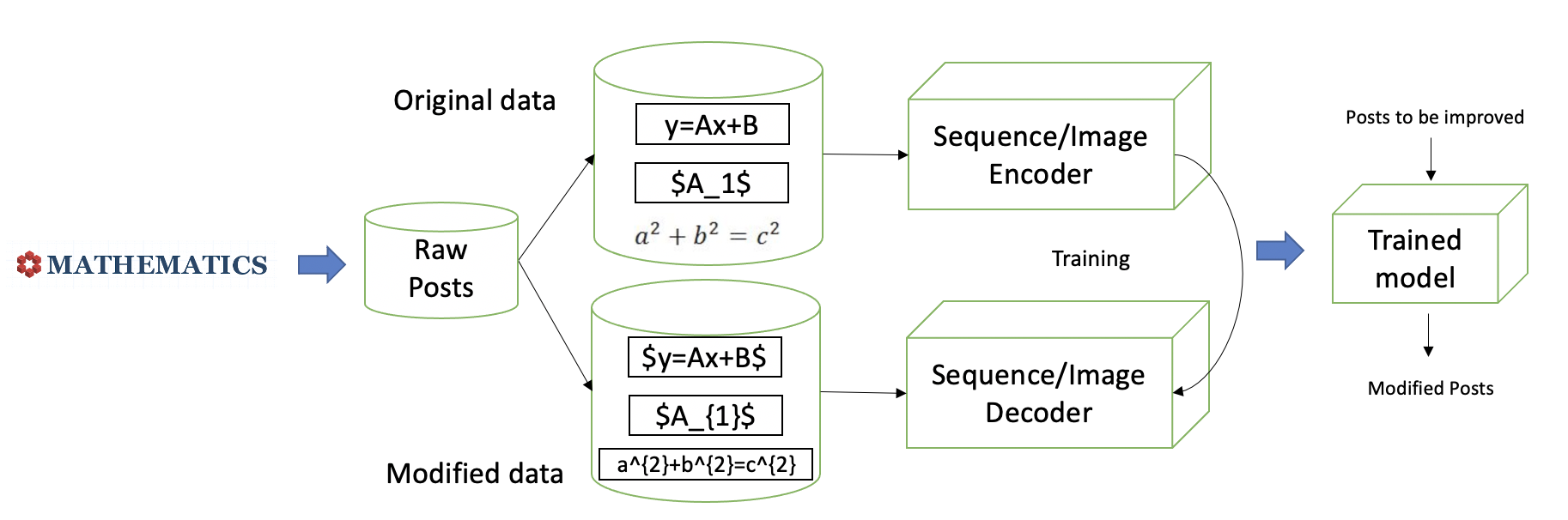}
	\caption{\revise{Workflow of our approach }
	}
	\label{fig:wkf}
\end{figure}

The workflow of our approach is shown in Fig \ref{fig:wkf}. 
Given three types of math-specific edit, we separate them into two tasks i.e., textual LaTeX edit (\textit{formula latexification}, \textit{LaTeX revision}) and visual LaTeX edit (\textit{screenshot transcription}).
Our approach first collected a large corpus of original-edited sentence pairs of modifying math formulas for subsequent textual LaTeX edit and synthesized a large corpus of image-formula pairs for model training for subsequent visual LaTeX edit (Section ~\ref{data_collection}). 
For textual LaTeX edits, our approach trained a transformer based model  on a large parallel corpus of original-edited sentence pairs (Section ~\ref{textual_edit}).
For visual LaTeX edits, our approach adopted an encoder-decoder model for converting the formula screenshot to LaTeX representation based on synthesized image-formula pairs (Section ~\ref{visual_edit}).
We specify the implementation details in Section~\ref{implement}. 
 

\subsection{Data collection}
\label{data_collection}
\subsubsection{For formula latexification and LaTeX revision tasks}
A post may have been edited several times.
Assume a post has $N$ versions, i.e., undergoing $N - 1$ post edits. For each post edit $i (1 \leq i \leq N - 1)$, we collect a pair of the original and edited content.
The original content is from the version $i$ of the post before the edit, and the edited content is from the version $i + 1$ of the post after the edit. 
\newrevise{
Then, we split the content into sentences.}

\newrevise{
We then align the sentence list $oList$ from the original content and the sentence list $eList$ from the edited content. 
For a sentence in the $oList$, if the similarity score of the most similar sentence in the $eList$ is above a threshold, the two sentences are aligned as a pair of original-edited sentences.
To calculate the similarity between one original sentence $l_o$ and edited sentence $l_e$, we calculate the Levenshtein distance~\cite{levenshtein1966binary} $dis(l_o, l_e)$ between two sentences.
The similarity threshold should be set to achieve a balanced precision and recall for sentence alignment. 
Therefore, we experimentally set the threshold at 0.9 in this work.
Note that some small edits may not influence readability a lot, however, the aggregating effect of several small issues in one post are not human-tolerable. 
Therefore, we still take them into consideration during data collection.}

We also filter out non-mathematical and noisy edits using the rule below: 
We remove the sentence pairs that do not include LaTeX formulas in original and edited sentences and sentence pairs in which the LaTeX formula does not change.
We also remove too long (char number $>256$) or too short (char number $<10$). 
In total, we collect 220,093 sentence pairs before March 1, 2020.

\subsubsection{For formula screenshot transcription task}
In Mathematics Stack Exchange, most users follow the guidelines by posting mathematics equations into LaTeX rather than as an image.
To collect the data for training our visual LaTeX edit model, we collected all LaTeX mathematics equations from \site and then rendered them to an image with automated scripts.
First, we extract  mathematics formulas using regular expressions i.e., extracting text within special annotations like ``\verb|begin\{equation\}(.*?)end\{equation\}|'' and ``\verb|\$([^\$]*?)\$|''.
By matching the raw posts content with regular expressions, we collected 5,505,098 raw LaTeX snippets.


Second, to filter out some noisy data, we removed the sequences without any mathematics characters such as ``+'', ``\verb|\frac|''.
We also removed duplicate LaTeX sequences and too long (char number $>256$) or too short (char number $<10$) LaTeX sequences.

\begin{figure}
	\centering
	\includegraphics[scale=0.2]{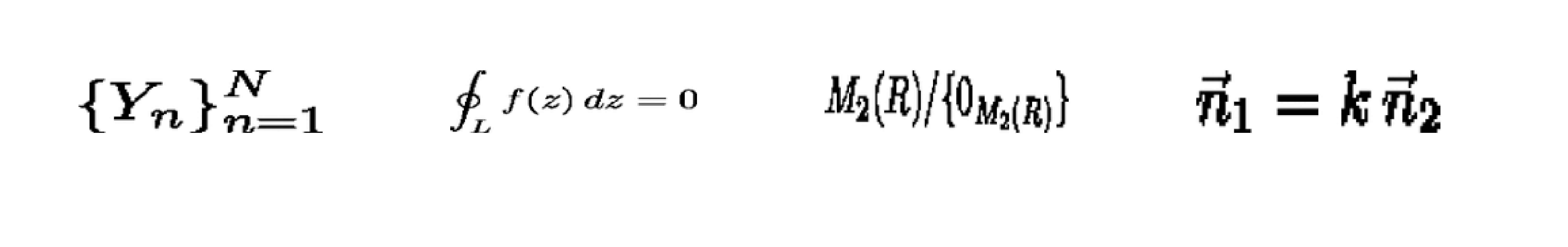}
	\caption{Examples of synthesized data}
	\label{fig:data_sample}
\end{figure}

Third, we tokenized the 2,068,744 remaining LaTeX sequences into separate words, especially the domain-specific tokens such as "$ \verb|\sum|$", "$ \verb|\frac|$" and "$ \verb|\sin|$" as one word.
To avoid ambiguity i.e., the same image can be generated by different LaTeX, we developed a LaTeX parser to keep the same LaTeX formatting.
For instance, we replace  ``$\verb|A^{c}_{2}|$'' with ``$\verb|A_{2}^{c}|$'', change ``$\verb|\over|$'' to ``$\verb|\frac{}{}|$'' and delete ``$\verb|\label{}|$''.
We converted these LaTeX formulas to images with $pdfLaTeX$ and excluded any formulas that failed to compile. 
Note that these synthesized images are different from the real-world screenshots.
The synthesized images are generated following the same rules while real-world screenshots differ greatly from users to users.
To bridge the gap between the GUI screenshots from the two resources, we applied an image augmentation method to transform the synthesized images to mimic those found in real-world images.
To do this, we randomly applied different sizes and resolution (i.e., DPI(dots per inch)) when rendering the synthesized images based on the collected LaTeX formula.
Some resulting formula screenshot examples by the image augmentation can be seen in Fig \ref{fig:data_sample}.

\subsection{Textual Latex Edit Recommendation}
\label{textual_edit}
\begin{figure}
	\centering
	\includegraphics[scale=0.35]{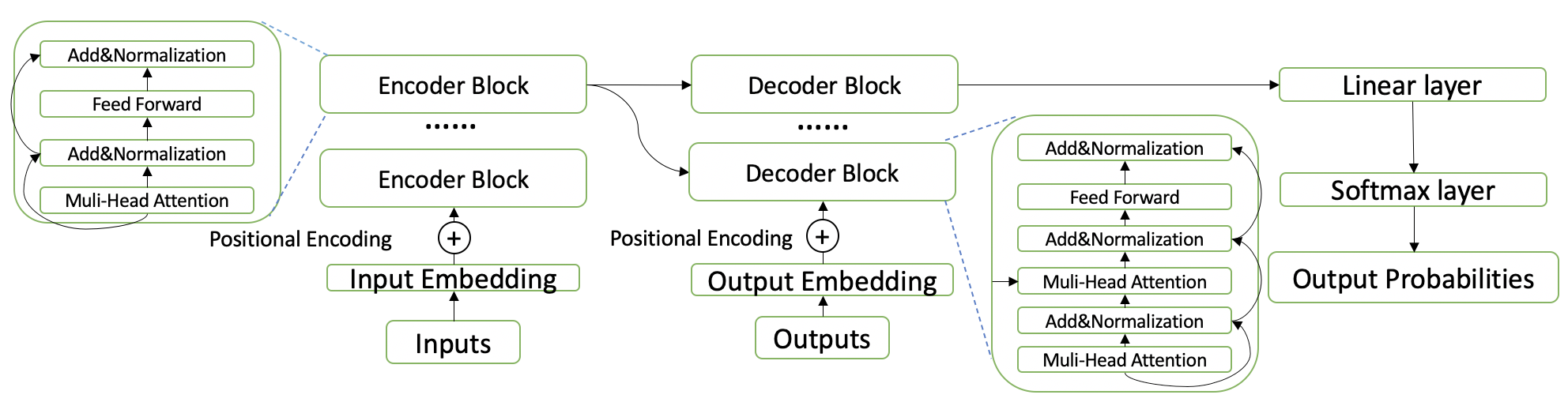}
	\caption{\revise{Textual LaTeX edit structure of {\tool}}}
	\label{fig:S_str}
\end{figure}


The textual LaTeX edit recommendation can be treated as a machine translation problem by treating the original post sentence as input and edited sentence as output.
Therefore, we adopt neural machine translation model~\cite{gao2019neural, wang2019domain} to learn the mapping from the source sentence to the target sentence.

\newrevise{
As shown in fig \ref{fig:S_str}, an attention-based transformer architecture is used for formula latexification and LaTeX revision.
Given the source word tokens $(x_1, ..., x_N )$, the goal is to predict the target word tokens $(y_1, ..., y_T )$.
The source word tokens are the original post sentence, and the predicted output word tokens are the edited post sentence.}
\revise{
The performance of deep learning models heavily depends on the quality of the training data.
In particular, our transformer model is sensitive to the input length i.e., the longer input sequence always results in worse performance as it is hard for the model to capture the long semantic especially for those long complex math formulas.
To mitigate that issue, we develop a normalization way to preprocess the input before feeding it to the model.}
Many sentences within our dataset is of both mathematical formula and natural-language part.
According to our observation, the non-math related content is rarely changed and math related content is always of special symbols like punctuations (e.g., ``\$+-=*''), numbers (e.g., 2, 3.14), variables (e.g., a, b, x, y) or commands in LaTex (e.g., $\log, \sin$).
Therefore, we manually construct a list of rules for detecting the potential mathematical content within the sentence, and replace the non-math part with the same placeholder symbol in the input.
To further shorten the length of input sequence, we also normalize the number with another placeholder symbol since it is rarely changed during the edit.
\newrevise{For example, the original sentence "{\small my first though was to factor by doing  ( 2  +  e $\widehat{}$ x  -  e $\widehat{}$ x )  /  ( e $\widehat{}$  (  - x )  + 1 )  but that negative in the denominator is not letting me solve the problem ?}" is preprocessed into "{\small \underline{COMMON\_WORDS}   ( 2  +  e $\widehat{}$ x  -  e $\widehat{}$ x )  /  ( e $\widehat{}$  (  - x )  + 1 )  \underline{COMMON\_WORDS}}", which highly shortens the length of the input.}

The normalized sentence pairs are used to train or test our transformer model.
Given an original sentence to be edited, the trained model will change it into an edited sentence.
The special symbols are then mapped back to the original domain-specific words in a post-processing step.
\newrevise{
For instance, the edited sentence "{\small \underline{COMMON\_WORDS}  \$ $\backslash$frac\{ 2  +  e $\widehat{}$ x  -  e $\widehat{}$ x \}\{ e $\widehat{}$  (  - x )  + 1\} \$   \underline{COMMON\_WORDS}}" is mapped back to "{\small my first though was to factor by doing  \$ $\backslash$frac\{ 2  +  e $\widehat{}$ x  -  e $\widehat{}$ x \}\{ e $\widehat{}$  (  - x )  + 1\} \$  but that negative in the denominator is not letting me solve the problem ?}".
}

\subsection{Visual Latex Edit Recommendation}
\label{visual_edit}
\begin{figure}
	\centering
	\includegraphics[scale=0.35]{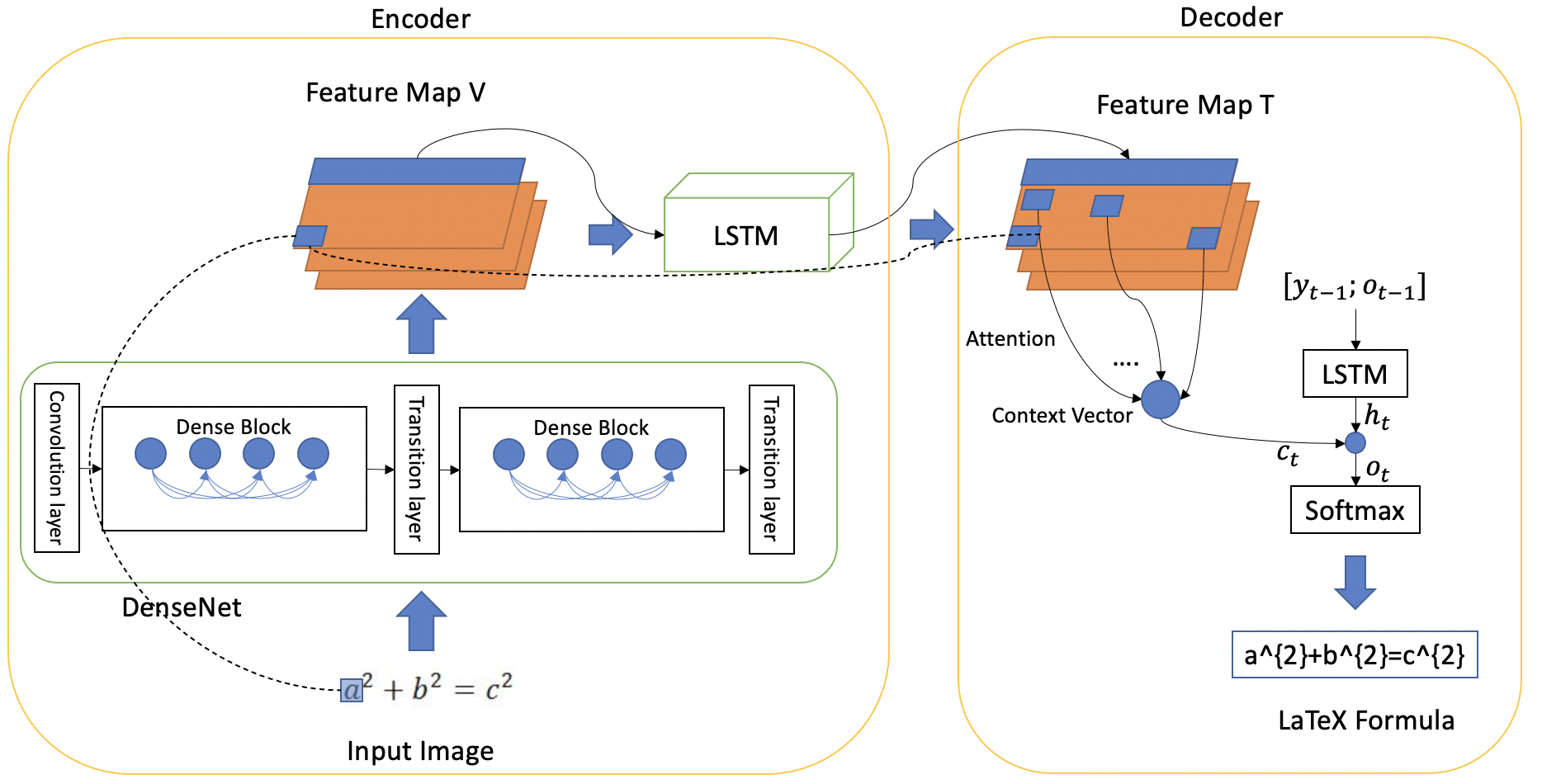}
	\caption{\revise{Visual LaTeX edit structure of {\tool}}}
	\label{fig:str}
\end{figure}

Our {\tool} tool applies an encoder-decoder structure for screenshot transcription, in which the encoder uses DenseNet, and the decoder uses LSTM with attention mechanism.
The overall structure is shown in Fig \ref{fig:str}.


\newrevise{
To get the feature maps of the input images, DenseNet is first used in the encoder to extract a feature map $V$. 
Unlike traditional Convolutional Neural Network (CNN), DenseNet connects each layer to every subsequent layer~\cite{huang2017densely,wang2019image}. 
The output features of DenseNet contain sequential order information. Thus we use another LSTM encoder to re-encode each row of DenseNet's output feature map. 
Based on the feature map $T$, we use LSTM and an attention mechanism~\cite{vaswani2017attention} as decoder to generate a sequence of predicted LaTeX tokens. 
}




\newrevise{
}

\label{inference}
\begin{figure}
	\centering
	\includegraphics[scale=0.35]{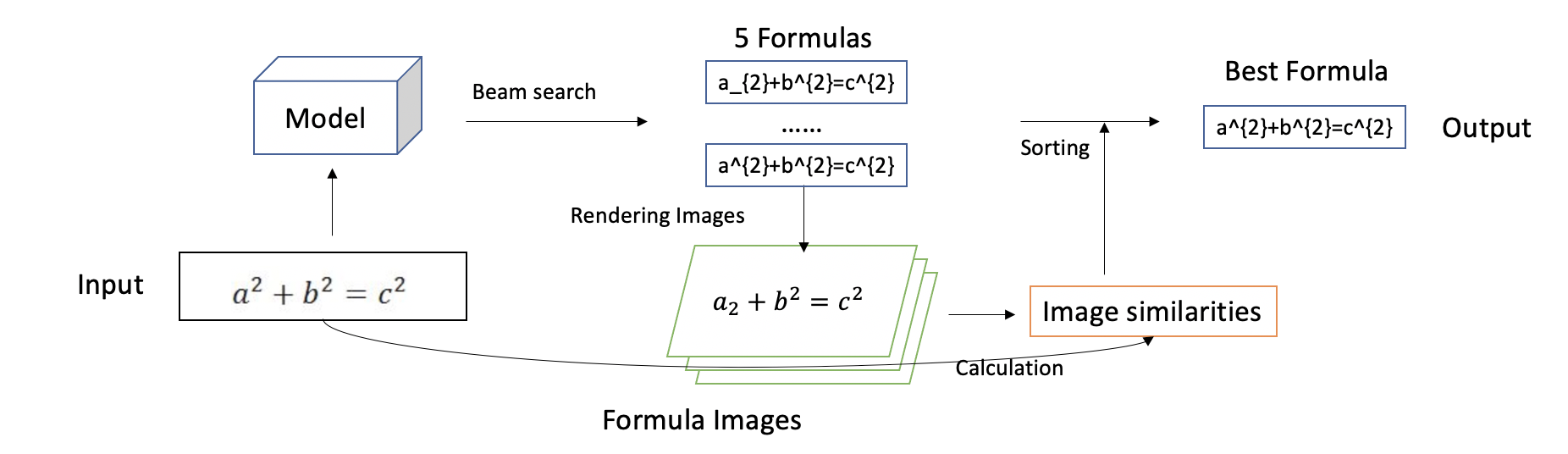}
	\caption{\revise{Overview of inference with image similarity}}
	\label{fig:infer}
\end{figure}
After training the visual LaTeX edit model, we use it to generate the LaTeX sequence $l$ for a mathematics screenshot $s$.
As seen in Fig~\ref{fig:infer}, we first adopt a beam search approach to select the top-5 candidates.
For each candidate, we render it into a formula image to compare the similarity of it with the original one and then select the most similar one as the generated result.

\newrevise{
Given the input screenshot, the generated LaTeX sequence should have the maximum log probability $p(l|s)$.
However, generating a global optimal LaTeX sequence has an immense search space.
Therefore, we adopt a beam search to expand only a limited set of the most promising nodes in the search space.}

After selecting the top-5 candidates with the highest probability, we further render them into mathematical formula images.
We compare these rendered images with the original input screenshot, and re-rank 5 candidates by image similarity.
To calculate image similarity, we firstly crop and resize these two images to same size. 
We then convert images to a binary image with pixel values as either 0 or 1, where 0 as black and 1 as white. 
For each column, we can convert a sequence of 0 and 1 into a value with binary system.
Let $a_i$ be the image array that is converted from original image $i$. 
Let $a_j$ be the image array that is converted from rendered image $j$.
We calculate the Levenshtein distance between $a_i$ and $a_j$ as $dis(a_i,a_j)$.
We can get the image similarity as $1-dis(a_i,a_j)/max(len(a_i),len(a_j))$, where $max(len(a_i),len(a_j))$ is the max length of $a_i$ and $a_j$.
Based on the image similarity of 5 candidates, we select the one with the highest similarity and its corresponding LaTeX sequence as the generated result of {\tool}.

\subsection{Implementation}
\label{implement}
{\tool} is based on Pytorch\footnote{\url{https://pytorch.org}} (textual LaTeX edit recommendation model) and Torch\footnote{\url{http://torch.ch/}} (visual LaTeX edit recommendation model) and all experiments were run on a 11GB NVidia TITAN Xp. 
The following settings are same for the two models. 
Minibatch stochastic gradient descent is used to optimize the parameters. 
The initial learning rate is set as 0.1. 
The training epoch is set as 25 and perplexity is used to select the best model during the validation step.
Once the validation score does not decrease, we halve the learning rate. 
Beam search is used during the test step, and the beam size is 5.
We also release the source code\footnote{\url{in https://github.com/astra1230/MathFormGen} } of this project.
\section{EVALUATING THE QUALITY OF \textit{\tool}}     
\label{results}
Our {\tool} tool aims to help \site Q\&A site post owners and editors to more effectively and collaboratively edit the math-related content in the post. 
The quality of the generated recommendation will hence greatly affect the utilization of {\tool} by the community. 

\subsection{Dataset}
For textual LaTeX edit recommendation, from 8,530,558 posts in Math Stack Exchange, 
we collected 219,420 original-edited sentence pairs about formula latexification and LaTeX revision. 
We randomly took 186,507 (85\%) of these sentence pairs as the training data, 10,971 (5\%) as the validation data and 21,942 (10\%) as the testing data to evaluate the quality of recommended edits by our tool.

For visual LaTeX edit recommendation,  we collected 2,068,744 LaTeX sequences.
Based on these extracted formulas, we generated 1,000,000 image-formula pairs. 
We randomly selected 900,000 (90\%) of these pairs as the training data, 50,000 (5\%) as the validation data to tune model hyperparamaters, and 50,000 (5\%)\footnote{We only use 5\% for testing as the image processing is much slower than text processing.} as the testing data ($test_{synthesize}$) to evaluate the quality of converted LaTeX sequences by {\tool}.
Apart from these synthesized data, we also collected the real-world post editing history where human editors have converted a posted math formula screenshot to LaTeX.
For each post edit, we compared the original and edited posts to check if some images in the original post are replaced by LaTeX sequence in the corresponding position of the edited post.
From 12,463 image-formula pairs that the images are manually converted to LaTeX sequences by editors, we randomly selected 100 of them as another testing set ($test_{real}$) to evaluate the performance of {\tool}.


\subsection{Baselines}
Apart from our own {\tool}, we selected other four methods as baselines for our comparisons.
For textual LaTeX edit recommendation,
the first baseline is a grammar error correction sequence to sequence (Seq2seq) model that contains a bidirectional RNN as an encoder and an attention-based decoder ~\cite{yuan2016grammatical}. 
The other baseline is the phrase-based Statistical Machine Translation (SMT) model~\cite{ortiz2005thot} specifically designed for sentence correction.
\revise{To check the influence of sentence normalization, we also take the derivative of our approach as the baseline i.e., our own {\tool} without sentence normalization.}
For all deep-learning models, we use the same training data to train the model.

For visual LaTeX edit recommendation,
the first baseline is the InftyReader, a commercial software mathematical expression recognition system. 
This tool is an OCR-based system, which combines symbol recognition and structural analysis phases \cite{SuzukiMasakazu2003IaiO-infty}.
The second baseline is the deep learning based model, WYGIWYS, which is specially designed to convert formula images to LaTeX sequences \cite{deng2016you}. 
\revise{We also add our approach derivative i.e., {\tool} without image similarity as another baseline.}


\subsection{Evaluation metrics}

\newrevise{
We adopt the BLEU (BiLingual Evaluation Understudy)~\cite{papineni-etal-2002-bleu} for evaluating the  quality of textual and visual LaTeX edit recommendation.
BLEU is an automatic evaluation metric widely used in machine translation studies.
It calculates the similarity of machine-generated translations and human-created reference translations (i.e., ground truth).
}

GLEU (Generalized Language Understanding Evaluation) is also used for evaluating textual LaTeX edit recommendation.
In the GEC field, recent released shared tasks have prompted the development of GLEU for evaluating GEC approaches  ~\cite{napoles2016gleu}. 
~\revise{
GLEU is a customized metric from the BLEU score which is widely used to evaluate the machine-translation quality. \cite{yuan2016grammatical}
It is independent of manual-annotation scheme and requires only reference sentences (without annotations of gold-standard edits).
Recent study shows that GLEU has the  correlation with human judgments of GEC quality and effort~\cite{napoles2016gleu}.}
Since it requires the input sequence, we do not use it for evaluating visual latex recommendation.

Apart from BLEU score, we also adopt the image similarity to check the quality of visual LaTeX edit recommendation by measuring the image similarity between the given screenshot and rendering image based on the generated Latex.
As the generated LaTeX can also be used to render the math image, we use the predicted LaTeX sequences and ground-truth LaTeX sequences to generate pair of images with the same resolution. 
We binarize the pair of images so that the pixel values are all 0 and 1, where 0 means black and 1 means white.
Then we convert the image into a one dimension array and remove the elements that only contain 1 value.
The following steps for image similarity is the same as what we use in Section\revise{~\ref{visual_edit}}.

\subsection{Evaluation Results}
We report the evaluation results of {\tool} from two aspects i.e., the performance of {\tool} in recommending edits to formula latexification and LaTeX revision, and how accurate {\tool} can be to generate the LaTeX sequence for a given math screenshot.

\subsubsection{Performance of textual LaTeX edit recommendation}

\begin{table}
	\caption{The Performance of different methods for textual LaTeX edit  models 
	}
	\small
	\begin{center}
		\begin{tabular}{l|c|c}
		    \hline
			Models &BLEU score&GLEU score\\
			\hline
			SMT&59.22&52.64\\
			Seq2seq&78.60&72.91\\
			\revise{{\tool} without sentence normalization}&80.41&74.69\\
			{\tool} &\textbf{82.30}&\textbf{76.57}\\	
			\hline
			
		\end{tabular}
		\label{tab:Seq_res}
	\end{center}
	\vspace{-2mm}
\end{table}

\begin{table}
	\caption{Examples of textual LaTeX edit  with different methods}
	\small
	\begin{center}
	\resizebox{\textwidth}{!}{

		\begin{tabular}{l|l|l|l|}
		\toprule

		Original sentence&Our model&Seq2seq&SMT\\

			\hline
			1 formula: y + py = px - 2p for which value ( s ) of p 1
			& formula: \underline{\$ y + py = px - 2p \$} for which value ( s ) of \underline{ \$ p \$ }1
			& formula: \underline{\$ y + py = px - 2p \$} for which value ( s ) of p 1
			& formula : \underline{\$ y + py = px - 2p \$} for which value ( s ) of p 1\\
		
			2  ' i ' is part of the ratio
			& \underline{ \$ i \$} is part of the ratio
			&  \underline{'\$ i \$'} is part of the ratio
			&  ' i ' is part of the ratio\\
			
			3 we have seen also that the primitives  f ( x , y )  

			&  we have seen also that the primitives  \underline{\$ f ( x , y )  \$ } 

			& we have seen also that the  \underline{starts \$ f ( x, y ) \$  }

			& we have seen also that the primitives   \underline{\$ f ( x , y )  \$}  \\
			
			4 can some one explain \$ f ( n )   =  10 * log ( n )  \$ 
			& can some one explain \underline{ \$ f ( n )   =  10  $\backslash cdot$ $\backslash$ log ( n )  \$}
			& can some one explain \underline{\$ f ( n )   =  10 *  $\backslash$ log ( n )  \$}
			& can some one explain  \underline{\$ f ( n )   =  10 * log ( n )  \$} \\
            
		\bottomrule

		\end{tabular}
		}
		\label{tab:Seq_example}
	\end{center}
	\vspace{-2mm}
\end{table}







\begin{table}
	\caption{The Performance of different methods with $test_{synthesize}$  and $test_{real}$ for visual LaTeX edit 
	}
	\small
	\begin{center}
		\begin{tabular}{l|c|c}
		    \hline
			Models &BLEU score&Image Similarity\\
			\hline
			INFTY($test_{synthesize}$)&67.72&50.21\\
			INFTY($test_{real}$)&47.24&43.12\\
			WYGIWYS($test_{synthesize}$)&89.92&90.22\\
			WYGIWYS($test_{real}$)&70.21&77.21\\
			\revise{{\tool}($test_{synthesize}$) without image similarity}&90.32&91.45\\
			\revise{{\tool}($test_{real}$) without image similarity}&73.21&79.36\\
			{\tool}($test_{synthesize}$)&\textbf{91.78}&\textbf{92.23}\\
			{\tool}($test_{real}$)&\textbf{74.71}&\textbf{81.61}\\
			\hline
		\end{tabular}
		\label{tab:Res}
	\end{center}
	\vspace{-2mm}
\end{table}

Table \ref{tab:Seq_res} presents the two metrics score of different methods for modifying post sentences.
Our {\tool} achieves the best overall result with the average BLEU score as 82.30, GLEU score as 76.57 which is 4.7\%, 5.0\% higher than Seq2seq model and 38.97\%, 45.46\% higher than SMT.
The improvement in the two scores by our model represents a significant improvement over the two baseline methods.
\revise{Note that in more detail, our model achieve 83.74 BLEU score, 77.41 GLEU score in formula latexification and 81.63 BLEU score, 76.03  GLEU score in LaTex revision.}

To qualitatively understand the strengths and weaknesses of different methods, we analyzed and compared the test results from three methods.
Table~\ref{tab:Seq_example} lists some representative examples in which {\tool} outperforms the two baseline methods.
Each row contains an original sentence and three edited sentences which are modified by different methods.
SMT can edit some domain-specific words (e.g., f(x,y) to \$f(x,y)\$ in 3rd example).
But it often preserves the original sentences that should be edited.
For instance, SMT fails to convert "p" into "\$ p \$" in 1st example, fail to convert "'i'" into "\$ i \$" in 2nd example and fail to convert "*" and "log" into "$\backslash$cdot" and "$\backslash$log" in 4th example.
Therefore, SMT does not work well for minor math-related changes in post edits because it cannot fully utilize the context information.

Seq2seq works better than SMT, but still fails to edit some tokens. 
For example, "'i'" is incorrectly converted to "'\$ i \$'" with an additional "'" in 2nd example and "primitives" is incorrectly converted into "starts" in 3rd example, "primitives" is incorrectly converted into "starts" in 3rd example, and it fails to convert "*" into "$\backslash$cdot" in 4th example.
The reason may be that Seq2seq incorrectly learns some biased knowledge from the training dataset and cannot handle the rare patterns. 
However, our {\tool} can avoid the incorrect modification in sentence tokens with sentence normalization and provide better results in all the examples.

\revise{The ablation study of our model without sentence normalization shows that the domain-specific normalization contributes 2.3\% and 2.5\% improvement than the vanilla model in BLEU and GLEU score.}
By analyzing low-quality recommendations by our model, we find four main reasons why our recommendation does not match the ground truth.
\newrevise{
First, some sentences are edited to add more information which is beyond the context of a sentence, such as editing  "{\small  i need to find weight for x and y }" to "{\small  I need to find weight  \underline{\$ w \_ 1 \$  and  \$ w \_ 2 \$ }}".}
Our current model considers only the local context of a sentence. 
To support such complicated edits, the broader context of the sentence (i.e., previous and subsequent sentences) need to be considered in the future.


Second, our model may provide better results compared with the ground truth.
\newrevise{
For example, our model edit the sentence "{\small if f $\circ$ f is differentiable,  then f $\circ$ f $\circ$ f is differentiable ?  }" to "{\small if \underline{\$ f $\backslash$ circ f} \$ is differentiable, then \underline{\$ f $\backslash$ circ f $\backslash$ circ f \$} is differentiable?}".
While the ground truth is "{\small if  f $\circ$  f  is differentiable, then \underline{\$ f $\circ$ f $\circ$ f \$} is differentiable?}".}
Compared with the ground truth, our model not only convert "$\circ$" to "$\backslash$ circ", but also correctly add "\$" to "f $\circ$  f".

Third, different users may have different opinions regarding what should or should not be edited.
\newrevise{
For example, some users will edit the sentence {\small "what i know so far is  \$ m =  ( y \_ 2 - y \_ 1 )  /  ( x \_ 2 - x \_ 1 )  \$"} to {\small "what I know so far is   \$ m =  \underline{$\backslash$ frac } \{ y \_ 2 - y \_ 1 \}  /  \{ x \_ 2 - x \_ 1 \} \$".}}
However, many other users will not do that.
Many revert-back edits we see when collecting original-edited sentences are the evidence of such different opinions. 
Different editing opinions often result in non-obvious editing patterns, which machine learning techniques cannot effectively encode.

Forth, the sentence length is a crucial factor influencing the performance of the model. 
Our model sometimes cannot fully correct all the errors in long LaTeX sequence.
\newrevise{
For example, {\small so  \$ k = 0  . 5  *  sqrt (  - 16 )   =  2i \$          \$ f \_ 1 ( x )   =  e $\widehat{}$  ( 2ix )   =  cos ( 2x )   +  isin ( 2x )  \$              \$ f \_ 2 ( x )   =  e $\widehat{}$  (  - 2ix )   =  cos ( 2x )   -  isin ( 2x )  \$ is the solution} is converted into 
{\small so  \$ k = 0  . 5  \underline{ $\backslash$sqrt}   (  - 16 )   =  2i \$          \$ f \_ 1 ( x )   =  e $\widehat{}$  ( 2ix )   =  \underline{ $\backslash$cos ( 2x )   +  i$\backslash$sin ( 2x )  \$}              \$ f \_ 2 ( x )   =  e $\widehat{}$  (  - 2ix )   =  \underline{ $\backslash$cos ( 2x )   -  i $\backslash$sin ( 2x )}  \$ is the solution}.
However, our model fails to convert {\small 0 . 5 * } to {\small 0 . 5 $\backslash$cdot },  {\small e $\widehat{}$  (  2ix )  to e $\widehat{}$  \{   2ix \}} and {\small e $\widehat{}$  (  -2ix )  to e $\widehat{}$  \{   -2ix \}}.}
To support such long sequence edits, splitting the long-length post sentence need to be considered in the future.

\subsubsection{Performance of visual LaTeX edit recommendation}
Table \ref{tab:Res} presents the two metrics score of different methods for converting formula images to LaTeX formula in the $test_{synthesize}$
dataset of 50,000 image-formula pairs and $test_{real}$ of 100 image-formula pairs. 

In $test_{synthesize}$, our {\tool} achieves the best overall result with the average BLEU score as 91.78 and image similarity as 92.23.
Due to the limitation of conventional image processing, INFTY gets the worst performance among all three models.
The Deep learning based model, WYGIWYS has much better performance than INFTY.
But {\tool} can still achieve 2.0\% and 2.2\% boost in BLEU score and image similarity. 
In $test_{real}$, all three models show slightly worse performance compared with $test_{synthesize}$, as it is a more challenging task.
{\tool} can still outperform the other two baselines with reasonably good performance in all metrics. 
\revise{Adding re-ranking based on the image similarity during inferring helps our model improve 1.6\% BLEU score in synthesized data and 2.8\% image similarity score in real data.}


\begin{figure}
	\centering
	\includegraphics[scale=0.37]{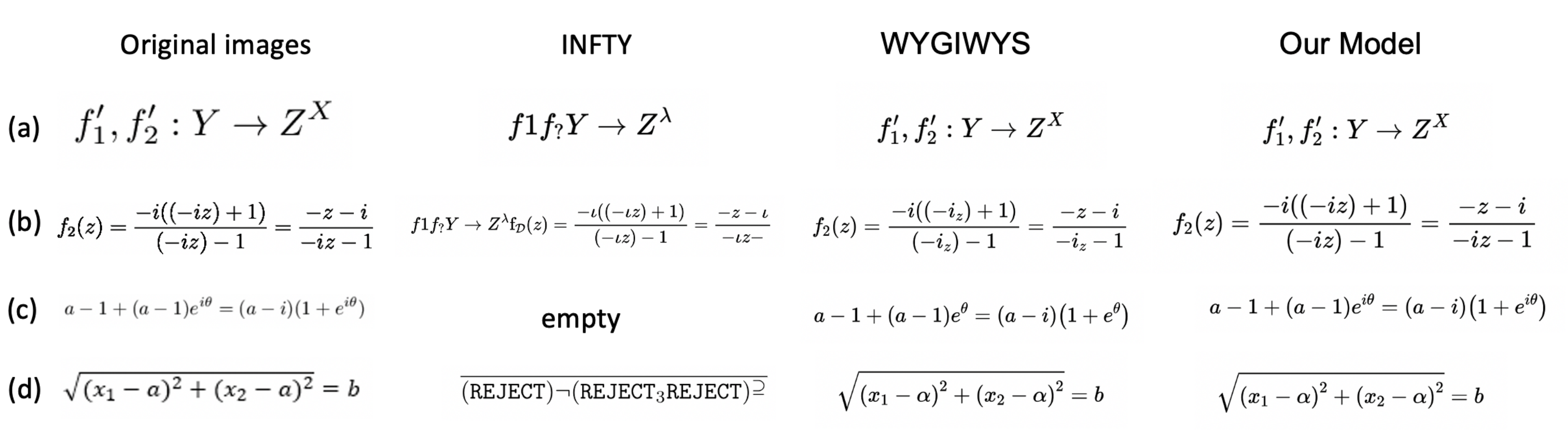}
	\caption{\newrevise{Examples of converted LaTeX formulas by different models}}
	\label{fig:compare}
\end{figure}

To qualitatively understand the strengths and weaknesses of different methods, we randomly selected 600 results from three methods for a detailed comparison.
Fig \ref{fig:compare} lists some representative examples in which {\tool}  outperforms the two baseline methods.  
Each row contains an original formula image and three formula images rendered by the predicted LaTeX sequences by three models.
INFTY does not work well for most cases as its rule-based methods do not scale well.
It first segments the characters in the image and then recognizes each character by comparing it with candidates in their database.
But there might be exceptions in each step and the errors in one step will be further amplified in the consecutive steps.
For example, it cannot segment the $f_1$ and $'$ from $f_{1}^{'}$ in Fig~\ref{fig:compare} (a).
The deep learning model, WYGIWYS behaves well in most cases, but it still makes mistakes especially for formulas with very fine-grained information.
For example, it incorrectly predicts ``$iz$'' as ``$i_z$'' in Fig~\ref{fig:compare}(b) and ``$e^{i\theta}$'' as ``$e^{\theta}$'' in Fig~\ref{fig:compare}(c), as these characters are too close to each other and relatively small compared with the whole screenshot.

\begin{figure}
	\centering
	\includegraphics[scale=0.3]{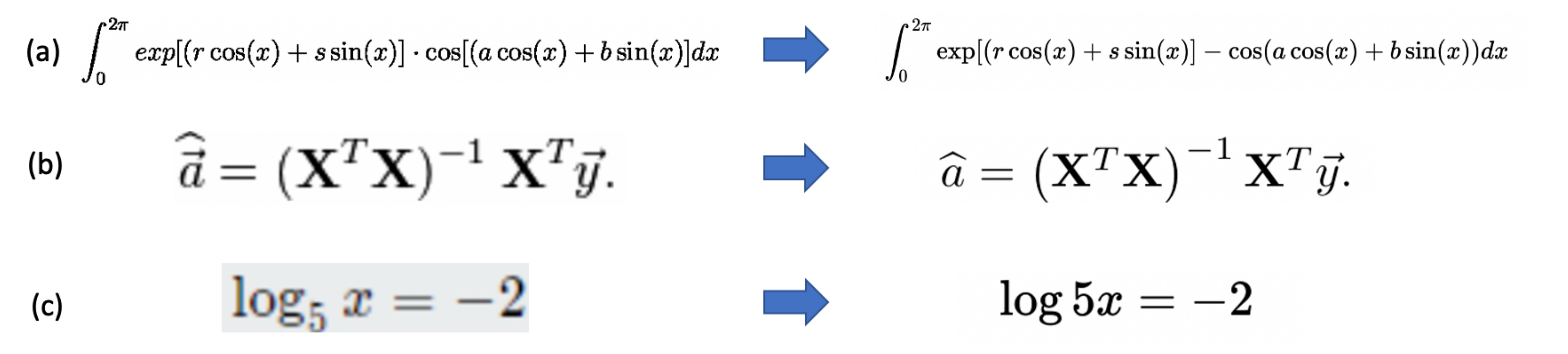}
	\caption{Examples of wrongly predicted LaTeX formulas by {\tool}}
	\label{fig:wrong}
\end{figure}

To analyze the low-quality recommendation by our model, we randomly selected some predictions which do not match the ground truth in both $test_{synthesize}$ and $test_{real}$.
According to our observation, we summarize three reasons for those erroneous predictions as Fig~\ref{fig:wrong}.

First, some input images have long or complex formulas, which makes it difficult to precisely predict the whole formula. 
For example, Fig \ref{fig:wrong}.(a) is one formula image whose formula has a long length and complex structure.
Although our predictions are right for most tokens, we still misclassify  ``$\cdot$'' as ``$-$'', ``$($'' and ``$)$'' as ``$[$'' and ``$]$''.
The longer the formula is, the more difficult we predict an exact match.

Second, some tokens in input formula images have some rare characters.
For example, the special combination of $\rightarrow$, $\widehat{}$ and $a$ as $\widehat{\Vec{a}}$ in Fig \ref{fig:wrong}.(b) is rarely used.
In the generated formula image, all other tokens except $\widehat{\Vec{a}}$ are correctly predicted, but $\widehat{\Vec{a}}$ is classified as $\widehat{a}$.
In our collected 2,068,744 math LaTeX sequences from Mathematics Stack Exchange, $\widehat{\Vec{}}$ only appears once, resulting in the miss in {\tool}.

Third, some input formula screenshots are too blurred or with some noisy background.
For example, Fig~\ref{fig:wrong}.(c) is a screenshot taken from a book with low resolution and gray background.
Most part of the generated LaTeX formula matches the ground truth, but there are still some minor errors like misclassifying "$\log_5$" as "$\log 5$".

\section{Real-world Evaluation of Edits}

\label{sec:usefulness}

Having established confidence in the quality of modified posts by {\tool}, we further investigated whether the recommended edits by {\tool} can actually assist post editors especially those with little editing experience or expertise to successfully improve the quality of mathematics posts in practice.
We conducted a user study and a field study  in this section to further evaluate the usefulness of our model for assisting domain-specific edits.

\subsection{Usefulness Evaluation}
\subsubsection{Procedures for User Study}
\revise{
We randomly selected 50 posts to manually check its needed edits.
We asked an experienced research staff (not involved in the study) to edit these posts and collected corresponding edits as the groundtruth.
Note that these posts are selected outside our training corpus (i.e., date after May 1, 2020) to avoid potential bias.
Among 50 posts, 14 of them require math-related edits and we further selected 6 needing different kinds of edits for this user study.}

\revise{We recruited ten PhD students and research assistants from our school. 
According to pre-study background survey, all participants knew \site and were familiar with using LaTeX. 
The study involves the two groups of five participants: the experimental group $P_1, P_2, P_3, P_4, P_5$ who do the post editing with our tool, and the control group $P_6, P_7, P_8, P_9, P_{10}$ who start from scratch. 
Each pair of participants $ ⟨P_x, P_{x+5}⟩ $ have comparable experience in using \site and Latex so that the experimental group has similar expertise to the control group in total. 
Note that we do not ask participants to edit half of the posts with our tool while the other half without assisting tools to avoid potential tool bias.}

\revise{We gave participants a detailed explanation of our tool for helping them understand the results.
Participants were required to edit posts with/without our tool and have up to 10 minutes for each post.
We recorded the time used to edit each post for every participant. 
\revise{
After each post editing, participants were asked to rate how satisfied they are with their edits in five-point Likert scale ~\cite{likert1932technique} (1: not satisfied at all and 5: highly satisfied). }
Compared with the groundtruth, we calculated the precision and recall for these post edits.}
\subsubsection{Results of User Study}
\begin{table}
	\caption{The comparison of the experiment and control group. * denotes $p < 0.01$ 
	}
	
	\small
	\begin{center}
		\begin{tabular}{ccc|cc|cc|cc}
			\hline
			\multirow{2}{*}{Post ID} & \multicolumn{2}{c}{Time cost (s)}&\multicolumn{2}{c}{Precision}&\multicolumn{2}{c}{Recall}&\multicolumn{2}{c}{Satisfactoriness}\\
			
			\cline{2-9}
			& CG & EG & CG & EG & CG & EG& CG & EG  \\
			\hline
			1&345.6&\textbf{56.4}&62.41&\textbf{90.24}&74.21&\textbf{91.62}&3.33&\textbf{4.50}\\
			2&373.6&\textbf{60.2}&70.21&\textbf{92.92}&70.53&\textbf{89.48}&3.50&\textbf{5.00}\\
			3&424.8&\textbf{70.2}&61.42&\textbf{94.75}&75.47&\textbf{93.70}&3.83&\textbf{4.83}\\
			4&483.4&\textbf{88.6}&71.23&\textbf{93.27}&78.64&\textbf{94.57}&3.50&\textbf{4.50}\\
			5&262.2&\textbf{60.6}&63.42&\textbf{92.88}&69.83&\textbf{90.29}&3.67&\textbf{4.83}\\
			6&382.4&\textbf{47.8}&60.74&\textbf{89.22}&71.42&\textbf{90.58}&3.67&\textbf{5.00}\\
			\hline
			Average&378.6& \textbf{64.0*}&68.70&\textbf{92.21*}&73.35&\textbf{91.70*}&3.58&\textbf{4.78*}\\
			\hline
			
		\end{tabular}
		\label{tab:time}
	\end{center}
	\vspace{-2mm}
\end{table}



Table \ref{tab:time} shows that participants in the experimental group spent less time (on average 64s versus 378.6s) in editing posts than the control group.
It indicates that our tool helps editors save 82.85\% editing time.
In fact, the average time of the control group is underestimated, as 2 participants failed to complete at least one post within 10 minutes, which means that they may need more time in the real editing.
In contrast, all participants in the experimental group finished all post editing within 3 minutes.
With the help of our \tool, the experimental group also obtained higher (34.22\% and 25.02\%) precision and recall score in terms of the edits.
The experimental group rated 80\% of their post edits as highly satisfactory (5 point), as opposed to 10\% highly satisfactory by the control group. 
On average, the satisfactoriness scores for the experiment and control group are 4.78 versus 3.58.

We carried out the Mann-Whitney U test~\cite{mwtest} (specifically designed for small samples) to understand the significance of the differences between the two groups.
It suggests that our tool can significantly help the experimental group edit posts faster ($p - value < 0.01$), with higher precision, recall and satisfactoriness score ($p - value < 0.01$).

\revise{
We believe that the better performance of the experimental group is due to the assistance of our {\tool} which gives participants a reliable starting point for editing. 
Guided by the recommended post edits, participants can easily identify the issues and finish the edit.
83.3\% recommendations provided by our tool are directly accepted by the editors.
Although there may be some mistakes in our tool's recommendation, participants can easily revise them. 
Without the help of edit recommendation, the control group have to read the post thoroughly and determine where issues are and solve these issues from scratch, which results in the longer edit time and less satisfactory edit results.
For example, some latexification edits within text like replacing x-y to $x-y$ are frequently missed in the control group.
The revision of complex latex formula such as $\langle x, x\rangle \cdot\langle y, y\rangle=|\langle x, y\rangle|^{2}+|x \times y|^{2}$ is more error-prone in control group.
}

\subsection{Real-world Assistance for Post Editors}

We also conducted a small-scale field study, in which the first author who has no experience in \site acted as a novice post editor. 
We randomly selected 600 posts (106 are with images) after May 1, 2020 which are never used in our training dataset to avoid potential bias.
Our model found that 112 posts need at least one LaTeX edit (including 95 posts with textual LaTeX edit, 22 posts with visual LaTeX edits).
Among those 112 posts, we selected 80 posts (60 with only textual Latex edits, 20 with only visual Latex edits) which is of reasonable size, and manageable with human effort to manually submit the post edits.
In Mathematics Stack Exchange, each question can have up to 5 tags to describe its topic. 
The 80 selected posts contain 203 tags in total (if the post is the answer, we took tags from its corresponding question) and 127 of these tags are unique.
This indicates that the 80 selected posts cover a diverse set of mathematical topics. 
In fact, these posts contain many mathematical terms that are beyond the expertise of the first author.
\revise{
Within 80 selected posts, 69 (86.25\%) of them just involve one edit while 11 (13.75\%) of them involve multiple edits.}
\begin{figure}
	\centering
	\includegraphics[scale=0.3]{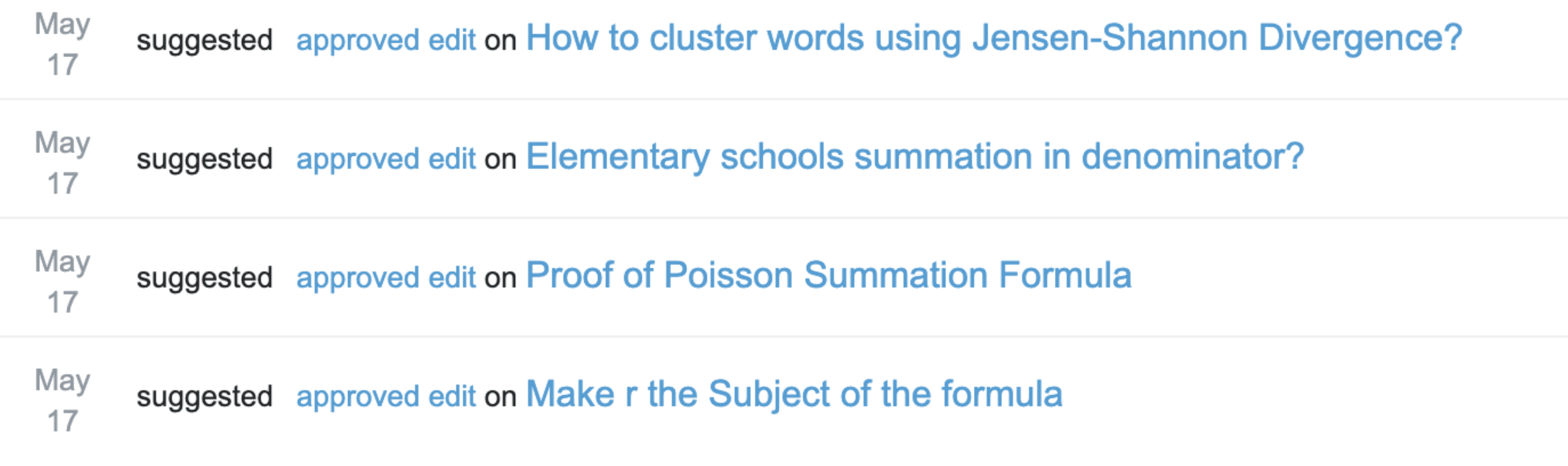}
	\caption{Example of {\tool} being used to assist post editors}
	\label{fig:web}
\end{figure}

\begin{figure}
	\centering
	\includegraphics[scale=0.2]{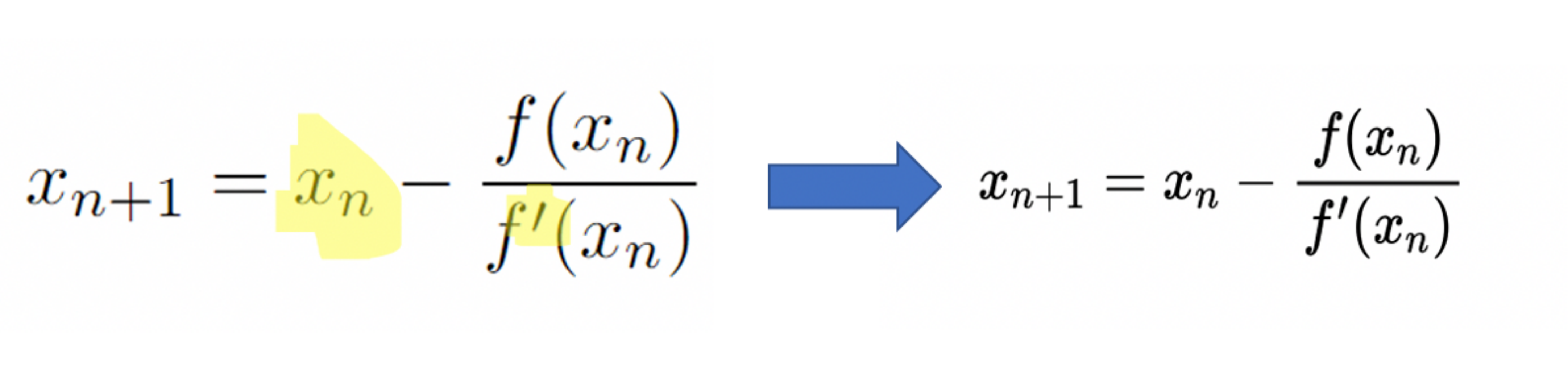}
	\caption{Example of rejected {\tool} generated revision}
	\label{fig:err2}
\end{figure}
\newrevise{
After proofreading the 80 posts edited by our {\tool}, he submitted the post edits to the community for approval.
For the 80 submitted post edits, 78  (97.5\%)  were accepted and 2 (2.5\%)  were rejected.
}
\revise{
And the two rejected ones are both single edits. 
The experiment results show that no matter there are single edits or multiple edits in one post, most of them are accepted by the community.}
Records of some edits are shown in Fig \ref{fig:web}.
For the two rejected edits, which are visual LaTeX edits, the original images have special notations in the formula.
Although {\tool} recommended the correct LaTeX sequences, the notations in the formulas images were missed.
Figure \ref{fig:err2} shows one example with yellow notation of some parts of the formula for illustrating the rejection.
Since our revised LaTeX sequence misses the notation which is important for this post, the revision was not accepted.
But for the 78 accepted post edits, the trusted contributors believed that they contained sufficient correct edits that significantly improved the post quality, and thus approved them.
This real-world acceptance of our {\tool}'s proposed edits demonstrates the usefulness of {\tool} in formula latexification, LaTeX revision and screenshot transcription for the \site Q\&A collaborative site.

\section{Discussion and Future Works}
In this section, we discuss the possible generalization of our {\tool}  approach for use to support collaboration on other Q\&A sites, and also its potential impact to improve accessibility of online mathematics-related information.

\subsection{Generalization of \textit{\tool}}
\label{sec:generalization}

This work examined collaborative editing patterns in Mathematics Stack Exchange, and we developed {\tool}, a deep learning-based approach for latexifing formula, revising latex and transcribing screenshots  to assist post owners and editors.
Note that our data analysis method and deep learning approach are totally data driven, and not tied to any specific collaborative editing or quality control process used by particular community Q\&A sites.
\revise{
In this work, we study only Mathematics Stack Exchange post edits. 
However, the input to our approach is essentially just a parallel corpus of original and edited text (see Figure \ref{fig:wkf}). }
Therefore, we would expect that our data analysis method and deep learning approach could be applied to other mathematics-related Q\&A sites.

\begin{figure}
	\centering
	\includegraphics[scale=0.38]{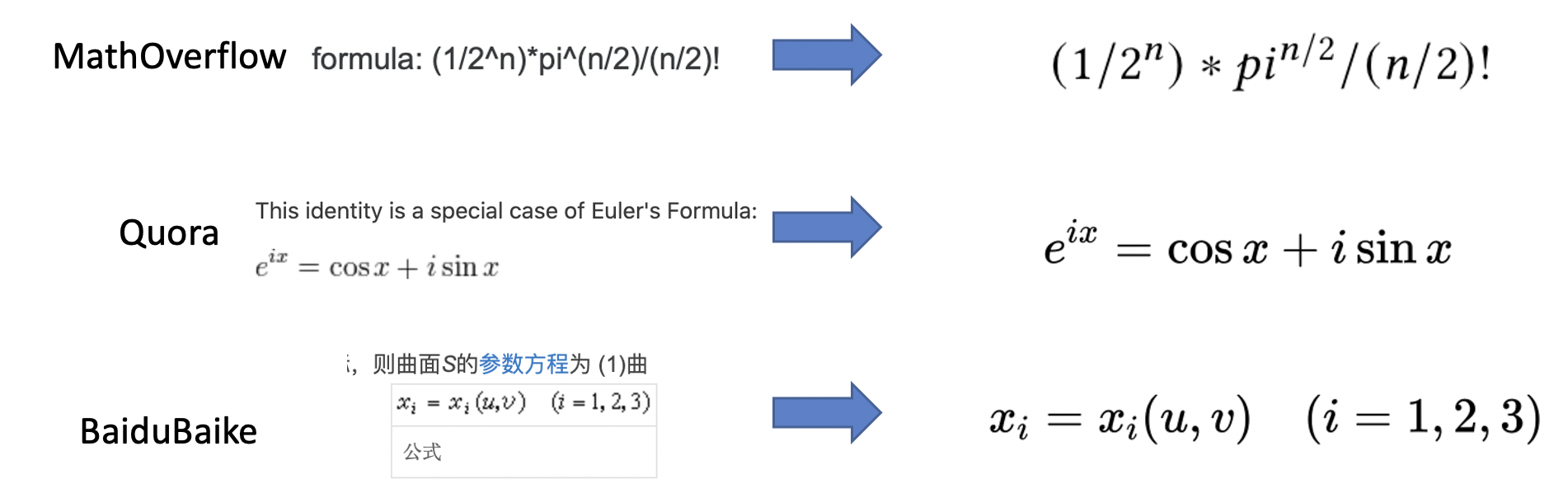}
	\caption{
	Example of {\tool} generated revision from other sites
	}
	\label{fig:baidu}
\end{figure}

There are also some other sites which contain mathematics formulas and also support collaborative editing, such as MathOverflow
\footnote{\url{https://mathoverflow.net/}} (math Q\&A site for professional mathematicians in Stack Exchange network), Quora\footnote{\url{https://www.quora.com/}} (largest general Q\&A site), BaiduBaike\footnote{\url{https://baike.baidu.com/}}(largest online encyclopedia created and edited by volunteers in China).
We randomly selected 15 posts that require formula latexification, latex revision or screenshot transcription from each site and edited them with our tool, as shown in Fig~\ref{fig:baidu}.
Since BaiduBaiKe requires certain reputation score for submission while we failed to reach that score, we only submitted our revisions in MathOverflow and Quora. 
We submitted our revisions to corresponding sites at 5 Oct, 2020.
MathOverflow accepted 15 edits, and Quora accepted 10 edits with the other 5 pending.
These results demonstrate the generalization of our {\tool} approach, and also the possibility of {\tool} to be adapted to use for these sites.


However, it is still an open question whether our approach can perform well in a large-scale, live deployment on mathematics Q\&A sites.
\newrevise{
Several issues require further investigation, such as: how to integrate our recommendation in the Q\&A and collaborative editing processes; can post owners and editors easily solve errors made in our generated LaTeX;  how does {\tool} usage impact the post owners and post editors' behavior; and how does the suggested edits of {\tool} lead to changes in people's perceptions of the math formula.}


\subsection{Implications of \textit{{\tool}} to Accessible Math}
Converting a posted mathematics formula screenshot to Latex brings many benefits to the site.
First, as shown in Fig~\ref{fig:compare}, transcribing the mathematics formula screenshot makes the site easier to read for everyone, as users can zoom in or out for a clearer view without blurring the formula.
Such readability can help other users better understand the question, resulting in higher possibility of responses.

Second, this machine-readability also makes the text easier to register and index by search-engine crawlers. 
This in turn makes the post easier to search for and easier to be found by other users, and this, therefore, makes the post more useful for a broader cross-section of internet users.

The screenshot transcribing feature would also make the questions more accessible to users with vision impairment, where the formula can be read by screen reader.
The users with visual impairments, ranging from difficulty reading all the way to complete blindness, need to be able to use the internet and study maths~\cite{AlajarmehNancy2012Dmma}.
When visiting the website, they mainly rely on assistive technology such as screen readers which can only read the text, rather than the image.
Using proper typesetting (including good use of Markdown and Mathjax) provides additional HTML syntax that those assistive technologies can use to give a more meaningful account of the content on the page.

Although {\tool} makes it easier for users to convert an image to LaTeX sequence, not all users will use it or follow the guidelines of Mathematics Stack Exchange.
Indeed, outside \site there are still many screenshots of mathematics formulas on different sites.
These inaccessible screenshots put a major barrier for users with vision impairment in learning and using mathematics online, resulting in degradation of their education equality.
{\tool} can help with converting a digital image into a LaTeX textual mathematics formula.
Physical books are another kind of important learning resources for those special students, our model has important potential to be adapted for mathematics formula recognition from physical or image-based electronic book pictures.

However, note that what we generate is only the LaTeX sequences, which are not very user friendly for most end users.
A simple division is written as ``\verb|\frac{numerator}{denominator}|'' as an example in LaTeX, and this is much more complicated than how we speak about the mathematics orally.
Therefore, in future work we want to develop a set of rules to further convert the generated LaTeX from {\tool} into the natural language of mathematics formula to make it more accessible for novice users.

\section{Conclusion}
In this paper, we carried out an empirical study of historical collaborative editing data on the Mathematics Stack Exchange.
Our results showed that collaborative editing is widely used in Mathematics Stack Exchange, which includes three domain-specific editing use cases as formula latexification, latex revision and screenshot transcription.
Due to the difficulty of the conversion, we designed {\tool}, a deep learning-based approach that automatically revises the math-related content in post. 
{\tool}'s recommendations can assist post owners and editors in improving dissemination of mathematics knowledge via the Q\&A site.
Our evaluation through use of large-scale  datasets demonstrates the quality of the edited LaTeX sequence by {\tool}. 
The edit recommendations from {\tool} for a selection of real-world posts were accepted by experienced users of Mathematics Stack Exchange, further showing the usefulness of our tool.
We discussed the potential benefits of our {\tool} post-edit recommendation approach for post owners/editors as well as different platforms.
However, deploying our approach on these sites may have complicated impacts on social process and collaborative editing, which need further study in the future.

\section*{ACKNOWLEDGMENTS}
Ma is supported by a Faculty of IT PhD scholarship.
Grundy and Khalajzadeh are supported by ARC Laureate Fellowship FL190100035.

\bibliographystyle{ACM-Reference-Format}
\bibliography{reference}


\begin{thebibliography}{57}


\ifx \showCODEN    \undefined \def \showCODEN     #1{\unskip}     \fi
\ifx \showDOI      \undefined \def \showDOI       #1{#1}\fi
\ifx \showISBNx    \undefined \def \showISBNx     #1{\unskip}     \fi
\ifx \showISBNxiii \undefined \def \showISBNxiii  #1{\unskip}     \fi
\ifx \showISSN     \undefined \def \showISSN      #1{\unskip}     \fi
\ifx \showLCCN     \undefined \def \showLCCN      #1{\unskip}     \fi
\ifx \shownote     \undefined \def \shownote      #1{#1}          \fi
\ifx \showarticletitle \undefined \def \showarticletitle #1{#1}   \fi
\ifx \showURL      \undefined \def \showURL       {\relax}        \fi
\providecommand\bibfield[2]{#2}
\providecommand\bibinfo[2]{#2}
\providecommand\natexlab[1]{#1}
\providecommand\showeprint[2][]{arXiv:#2}

\bibitem[\protect\citeauthoryear{??}{wik}{2020}]%
        {wiki:how}
 \bibinfo{year}{2020}\natexlab{}.
\newblock \bibinfo{title}{About wikiHow}.
\newblock
  \bibinfo{howpublished}{\url{https://www.wikihow.com/wikiHow:About-wikiHow}}.
\newblock
\newblock
\shownote{Accessed May 26, 2020.}


\bibitem[\protect\citeauthoryear{??}{hel}{2020a}]%
        {help1}
 \bibinfo{year}{2020}\natexlab{a}.
\newblock \bibinfo{title}{How do I write a good answer?}
\newblock
  \bibinfo{howpublished}{\url{https://math.stackexchange.com/help/how-to-answer}}.
\newblock
\newblock
\shownote{Accessed May 26, 2020.}


\bibitem[\protect\citeauthoryear{??}{hel}{2020b}]%
        {help3}
 \bibinfo{year}{2020}\natexlab{b}.
\newblock \bibinfo{title}{How to Ask}.
\newblock
  \bibinfo{howpublished}{\url{https://math.stackexchange.com/questions/ask/advice}}.
\newblock
\newblock
\shownote{Accessed May 26, 2020.}


\bibitem[\protect\citeauthoryear{??}{mat}{2020a}]%
        {math_latex}
 \bibinfo{year}{2020}\natexlab{a}.
\newblock \bibinfo{title}{MathJax basic tutorial and quick reference}.
\newblock
  \bibinfo{howpublished}{\url{https://math.meta.stackexchange.com/questions/5020/mathjax-basic-tutorial-and-quick-reference}}.
\newblock
\newblock
\shownote{Accessed Oct 16, 2020.}


\bibitem[\protect\citeauthoryear{??}{Wik}{2020}]%
        {Wiki}
 \bibinfo{year}{2020}\natexlab{}.
\newblock \bibinfo{title}{The Objective Revision Evaluation Service}.
\newblock \bibinfo{howpublished}{\url{https://ores.wikimedia.org/}}.
\newblock
\newblock
\shownote{Accessed May 25, 2020.}


\bibitem[\protect\citeauthoryear{??}{mat}{2020b}]%
        {math2}
 \bibinfo{year}{2020}\natexlab{b}.
\newblock \bibinfo{title}{Stack Exchange}.
\newblock \bibinfo{howpublished}{\url{https://stackexchange.com/sites##}}.
\newblock
\newblock
\shownote{Accessed May 28, 2020.}


\bibitem[\protect\citeauthoryear{??}{mat}{2020c}]%
        {math}
 \bibinfo{year}{2020}\natexlab{c}.
\newblock \bibinfo{title}{Welcome to Mathematics Stack Exchange}.
\newblock \bibinfo{howpublished}{\url{https://math.stackexchange.com/tour}}.
\newblock
\newblock
\shownote{Accessed May 26, 2020.}


\bibitem[\protect\citeauthoryear{??}{hel}{2020c}]%
        {help2}
 \bibinfo{year}{2020}\natexlab{c}.
\newblock \bibinfo{title}{Why can people edit my posts? How does editing work?}
\newblock
  \bibinfo{howpublished}{\url{https://math.stackexchange.com/help/editing}}.
\newblock
\newblock
\shownote{Accessed May 26, 2020.}


\bibitem[\protect\citeauthoryear{Alajarmeh}{Alajarmeh}{2012}]%
        {AlajarmehNancy2012Dmma}
\bibfield{author}{\bibinfo{person}{Nancy Alajarmeh}.}
  \bibinfo{year}{2012}\natexlab{}.
\newblock \showarticletitle{Doing math: mathematics accessibility issues}. In
  \bibinfo{booktitle}{\emph{Proceedings of the International Cross-Disciplinary
  Conference on web accessibility}} \emph{(\bibinfo{series}{W4A '12})}.
  \bibinfo{publisher}{ACM}, \bibinfo{pages}{1--2}.
\newblock
\showISBNx{9781450310192}


\bibitem[\protect\citeauthoryear{Blei, Ng, and Jordan}{Blei
  et~al\mbox{.}}{2003}]%
        {blei2003latent-lda}
\bibfield{author}{\bibinfo{person}{David~M Blei}, \bibinfo{person}{Andrew~Y
  Ng}, {and} \bibinfo{person}{Michael~I Jordan}.}
  \bibinfo{year}{2003}\natexlab{}.
\newblock \showarticletitle{Latent dirichlet allocation}.
\newblock \bibinfo{journal}{\emph{Journal of machine Learning research}}
  \bibinfo{volume}{3}, \bibinfo{number}{Jan} (\bibinfo{year}{2003}),
  \bibinfo{pages}{993--1022}.
\newblock


\bibitem[\protect\citeauthoryear{Cao, Chen, Baltes, Treude, and Chen}{Cao
  et~al\mbox{.}}{2021}]%
        {cao2021automated}
\bibfield{author}{\bibinfo{person}{Kaibo Cao}, \bibinfo{person}{Chunyang Chen},
  \bibinfo{person}{Sebastian Baltes}, \bibinfo{person}{Christoph Treude}, {and}
  \bibinfo{person}{Xiang Chen}.} \bibinfo{year}{2021}\natexlab{}.
\newblock \showarticletitle{Automated Query Reformulation for Efficient Search
  based on Query Logs From Stack Overflow}. In \bibinfo{booktitle}{\emph{2021
  IEEE/ACM 43rd International Conference on Software Engineering (ICSE)}}.
  IEEE, \bibinfo{pages}{1273--1285}.
\newblock


\bibitem[\protect\citeauthoryear{Cervone}{Cervone}{2012}]%
        {CervoneDavide2012MAPf}
\bibfield{author}{\bibinfo{person}{Davide Cervone}.}
  \bibinfo{year}{2012}\natexlab{}.
\newblock \showarticletitle{MathJax: A Platform for Mathematics on the Web}.
\newblock \bibinfo{journal}{\emph{Notices of the American Mathematical
  Society}} \bibinfo{volume}{59}, \bibinfo{number}{02} (\bibinfo{year}{2012}),
  \bibinfo{pages}{1}.
\newblock
\showISSN{0002-9920}


\bibitem[\protect\citeauthoryear{Chen, Chen, Sun, Xing, and Li}{Chen
  et~al\mbox{.}}{2018a}]%
        {chen2018data}
\bibfield{author}{\bibinfo{person}{Chunyang Chen}, \bibinfo{person}{Xi Chen},
  \bibinfo{person}{Jiamou Sun}, \bibinfo{person}{Zhenchang Xing}, {and}
  \bibinfo{person}{Guoqiang Li}.} \bibinfo{year}{2018}\natexlab{a}.
\newblock \showarticletitle{Data-driven proactive policy assurance of post
  quality in community q\&a sites}.
\newblock \bibinfo{journal}{\emph{Proceedings of the ACM on human-computer
  interaction}} \bibinfo{volume}{2}, \bibinfo{number}{CSCW}
  (\bibinfo{year}{2018}), \bibinfo{pages}{1--22}.
\newblock


\bibitem[\protect\citeauthoryear{Chen, Su, Meng, Xing, and Liu}{Chen
  et~al\mbox{.}}{2018b}]%
        {chen2018ui}
\bibfield{author}{\bibinfo{person}{Chunyang Chen}, \bibinfo{person}{Ting Su},
  \bibinfo{person}{Guozhu Meng}, \bibinfo{person}{Zhenchang Xing}, {and}
  \bibinfo{person}{Yang Liu}.} \bibinfo{year}{2018}\natexlab{b}.
\newblock \showarticletitle{From ui design image to gui skeleton: a neural
  machine translator to bootstrap mobile gui implementation}. In
  \bibinfo{booktitle}{\emph{Proceedings of the 40th International Conference on
  Software Engineering}}. \bibinfo{pages}{665--676}.
\newblock


\bibitem[\protect\citeauthoryear{Chen and Xing}{Chen and Xing}{2016}]%
        {chen2016mining}
\bibfield{author}{\bibinfo{person}{Chunyang Chen} {and}
  \bibinfo{person}{Zhenchang Xing}.} \bibinfo{year}{2016}\natexlab{}.
\newblock \showarticletitle{Mining technology landscape from stack overflow}.
  In \bibinfo{booktitle}{\emph{Proceedings of the 10th ACM/IEEE International
  Symposium on Empirical Software Engineering and Measurement}}.
  \bibinfo{pages}{1--10}.
\newblock


\bibitem[\protect\citeauthoryear{Chen, Xing, and Han}{Chen
  et~al\mbox{.}}{2016}]%
        {chen2016techland}
\bibfield{author}{\bibinfo{person}{Chunyang Chen}, \bibinfo{person}{Zhenchang
  Xing}, {and} \bibinfo{person}{Lei Han}.} \bibinfo{year}{2016}\natexlab{}.
\newblock \showarticletitle{Techland: Assisting technology landscape inquiries
  with insights from stack overflow}. In \bibinfo{booktitle}{\emph{2016 IEEE
  International Conference on Software Maintenance and Evolution (ICSME)}}.
  IEEE, \bibinfo{pages}{356--366}.
\newblock


\bibitem[\protect\citeauthoryear{Chen, Xing, and Liu}{Chen
  et~al\mbox{.}}{2017a}]%
        {chen2017community}
\bibfield{author}{\bibinfo{person}{Chunyang Chen}, \bibinfo{person}{Zhenchang
  Xing}, {and} \bibinfo{person}{Yang Liu}.} \bibinfo{year}{2017}\natexlab{a}.
\newblock \showarticletitle{By the community \& for the community: a deep
  learning approach to assist collaborative editing in q\&a sites}.
\newblock \bibinfo{journal}{\emph{Proceedings of the ACM on Human-Computer
  Interaction}} \bibinfo{volume}{1}, \bibinfo{number}{CSCW}
  (\bibinfo{year}{2017}), \bibinfo{pages}{1--21}.
\newblock


\bibitem[\protect\citeauthoryear{Chen, Xing, and Wang}{Chen
  et~al\mbox{.}}{2017b}]%
        {chen2017unsupervised}
\bibfield{author}{\bibinfo{person}{Chunyang Chen}, \bibinfo{person}{Zhenchang
  Xing}, {and} \bibinfo{person}{Ximing Wang}.}
  \bibinfo{year}{2017}\natexlab{b}.
\newblock \showarticletitle{Unsupervised software-specific morphological forms
  inference from informal discussions}. In \bibinfo{booktitle}{\emph{2017
  IEEE/ACM 39th International Conference on Software Engineering (ICSE)}}.
  IEEE, \bibinfo{pages}{450--461}.
\newblock


\bibitem[\protect\citeauthoryear{Chen, Chen, Xing, Xu, Zhut, Li, and Wang}{Chen
  et~al\mbox{.}}{2020}]%
        {chen2020unblind}
\bibfield{author}{\bibinfo{person}{Jieshan Chen}, \bibinfo{person}{Chunyang
  Chen}, \bibinfo{person}{Zhenchang Xing}, \bibinfo{person}{Xiwei Xu},
  \bibinfo{person}{Liming Zhut}, \bibinfo{person}{Guoqiang Li}, {and}
  \bibinfo{person}{Jinshui Wang}.} \bibinfo{year}{2020}\natexlab{}.
\newblock \showarticletitle{Unblind your apps: Predicting natural-language
  labels for mobile gui components by deep learning}. In
  \bibinfo{booktitle}{\emph{2020 IEEE/ACM 42nd International Conference on
  Software Engineering (ICSE)}}. IEEE, \bibinfo{pages}{322--334}.
\newblock


\bibitem[\protect\citeauthoryear{Chen, Chen, Zhang, and Xing}{Chen
  et~al\mbox{.}}{2019}]%
        {chen2019sethesaurus}
\bibfield{author}{\bibinfo{person}{Xiang Chen}, \bibinfo{person}{Chunyang
  Chen}, \bibinfo{person}{Dun Zhang}, {and} \bibinfo{person}{Zhenchang Xing}.}
  \bibinfo{year}{2019}\natexlab{}.
\newblock \showarticletitle{Sethesaurus: Wordnet in software engineering}.
\newblock \bibinfo{journal}{\emph{IEEE Transactions on Software Engineering}}
  (\bibinfo{year}{2019}).
\newblock


\bibitem[\protect\citeauthoryear{Choi and Tausczik}{Choi and Tausczik}{2018}]%
        {choi2018will}
\bibfield{author}{\bibinfo{person}{Joohee Choi} {and} \bibinfo{person}{Yla
  Tausczik}.} \bibinfo{year}{2018}\natexlab{}.
\newblock \showarticletitle{Will Too Many Editors Spoil The Tag? Conflicts and
  Alignment in Q\&A Categorization}.
\newblock \bibinfo{journal}{\emph{Proceedings of the ACM on Human-Computer
  Interaction}} \bibinfo{volume}{2}, \bibinfo{number}{CSCW}
  (\bibinfo{year}{2018}), \bibinfo{pages}{1--19}.
\newblock


\bibitem[\protect\citeauthoryear{Chollampatt and Ng}{Chollampatt and
  Ng}{2018}]%
        {chollampatt-ng-2018-neural}
\bibfield{author}{\bibinfo{person}{Shamil Chollampatt} {and}
  \bibinfo{person}{Hwee~Tou Ng}.} \bibinfo{year}{2018}\natexlab{}.
\newblock \showarticletitle{Neural Quality Estimation of Grammatical Error
  Correction}. In \bibinfo{booktitle}{\emph{Proceedings of the 2018 Conference
  on Empirical Methods in Natural Language Processing}}.
  \bibinfo{publisher}{Association for Computational Linguistics},
  \bibinfo{address}{Brussels, Belgium}, \bibinfo{pages}{2528--2539}.
\newblock
\urldef\tempurl%
\url{https://doi.org/10.18653/v1/D18-1274}
\showDOI{\tempurl}


\bibitem[\protect\citeauthoryear{Danesi}{Danesi}{2016}]%
        {DanesiMarcel.author2016LaTM}
\bibfield{author}{\bibinfo{person}{Marcel.~author Danesi}.}
  \bibinfo{year}{2016}\natexlab{}.
\newblock \bibinfo{booktitle}{\emph{Learning and Teaching Mathematics in The
  Global Village Math Education in the Digital Age} (\bibinfo{edition}{1st ed.
  2016.} ed.)}.
\newblock


\bibitem[\protect\citeauthoryear{Deng, Kanervisto, and Rush}{Deng
  et~al\mbox{.}}{2016}]%
        {deng2016you}
\bibfield{author}{\bibinfo{person}{Yuntian Deng}, \bibinfo{person}{Anssi
  Kanervisto}, {and} \bibinfo{person}{Alexander~M Rush}.}
  \bibinfo{year}{2016}\natexlab{}.
\newblock \showarticletitle{What you get is what you see: A visual markup
  decompiler}.
\newblock \bibinfo{journal}{\emph{arXiv preprint arXiv:1609.04938}}
  \bibinfo{volume}{10} (\bibinfo{year}{2016}), \bibinfo{pages}{32--37}.
\newblock


\bibitem[\protect\citeauthoryear{Fay and Proschan}{Fay and Proschan}{2010}]%
        {mwtest}
\bibfield{author}{\bibinfo{person}{Michael~P Fay} {and}
  \bibinfo{person}{Michael~A Proschan}.} \bibinfo{year}{2010}\natexlab{}.
\newblock \showarticletitle{Wilcoxon-Mann-Whitney or t-test? On assumptions for
  hypothesis tests and multiple interpretations of decision rules}.
\newblock \bibinfo{journal}{\emph{Statistics surveys}}  \bibinfo{volume}{4}
  (\bibinfo{year}{2010}), \bibinfo{pages}{1—39}.
\newblock
\showISSN{1935-7516}
\urldef\tempurl%
\url{https://doi.org/10.1214/09-ss051}
\showDOI{\tempurl}


\bibitem[\protect\citeauthoryear{Ford, Lustig, Banks, and Parnin}{Ford
  et~al\mbox{.}}{2018}]%
        {ford2018we}
\bibfield{author}{\bibinfo{person}{Denae Ford}, \bibinfo{person}{Kristina
  Lustig}, \bibinfo{person}{Jeremy Banks}, {and} \bibinfo{person}{Chris
  Parnin}.} \bibinfo{year}{2018}\natexlab{}.
\newblock \showarticletitle{" We Don't Do That Here" How Collaborative Editing
  with Mentors Improves Engagement in Social Q\&A Communities}. In
  \bibinfo{booktitle}{\emph{Proceedings of the 2018 CHI conference on human
  factors in computing systems}}. \bibinfo{pages}{1--12}.
\newblock


\bibitem[\protect\citeauthoryear{Gao, Chen, Xing, Ma, Song, and Lin}{Gao
  et~al\mbox{.}}{2019}]%
        {gao2019neural}
\bibfield{author}{\bibinfo{person}{Sa Gao}, \bibinfo{person}{Chunyang Chen},
  \bibinfo{person}{Zhenchang Xing}, \bibinfo{person}{Yukun Ma},
  \bibinfo{person}{Wen Song}, {and} \bibinfo{person}{Shang-Wei Lin}.}
  \bibinfo{year}{2019}\natexlab{}.
\newblock \showarticletitle{A neural model for method name generation from
  functional description}. In \bibinfo{booktitle}{\emph{2019 IEEE 26th
  International Conference on Software Analysis, Evolution and Reengineering
  (SANER)}}. IEEE, \bibinfo{pages}{414--421}.
\newblock


\bibitem[\protect\citeauthoryear{Garain, Chaudhuri, and Chaudhuri}{Garain
  et~al\mbox{.}}{2004}]%
        {garain2004identification}
\bibfield{author}{\bibinfo{person}{Utpal Garain}, \bibinfo{person}{BB
  Chaudhuri}, {and} \bibinfo{person}{Adrish~Ray Chaudhuri}.}
  \bibinfo{year}{2004}\natexlab{}.
\newblock \showarticletitle{Identification of embedded mathematical expressions
  in scanned documents}. In \bibinfo{booktitle}{\emph{Proceedings of the 17th
  International Conference on Pattern Recognition, 2004. ICPR 2004.}},
  Vol.~\bibinfo{volume}{1}. IEEE, \bibinfo{pages}{384--387}.
\newblock


\bibitem[\protect\citeauthoryear{Grundkiewicz and Junczys-Dowmunt}{Grundkiewicz
  and Junczys-Dowmunt}{2018}]%
        {grundkiewicz-junczys-dowmunt-2018-near}
\bibfield{author}{\bibinfo{person}{Roman Grundkiewicz} {and}
  \bibinfo{person}{Marcin Junczys-Dowmunt}.} \bibinfo{year}{2018}\natexlab{}.
\newblock \showarticletitle{Near Human-Level Performance in Grammatical Error
  Correction with Hybrid Machine Translation}. In
  \bibinfo{booktitle}{\emph{Proceedings of the 2018 Conference of the North
  {A}merican Chapter of the Association for Computational Linguistics: Human
  Language Technologies, Volume 2 (Short Papers)}}.
  \bibinfo{publisher}{Association for Computational Linguistics},
  \bibinfo{address}{New Orleans, Louisiana}, \bibinfo{pages}{284--290}.
\newblock
\urldef\tempurl%
\url{https://doi.org/10.18653/v1/N18-2046}
\showDOI{\tempurl}


\bibitem[\protect\citeauthoryear{Hochreiter and Schmidhuber}{Hochreiter and
  Schmidhuber}{1997}]%
        {hochreiter1997long-lstm}
\bibfield{author}{\bibinfo{person}{Sepp Hochreiter} {and}
  \bibinfo{person}{J{\"u}rgen Schmidhuber}.} \bibinfo{year}{1997}\natexlab{}.
\newblock \showarticletitle{Long short-term memory}.
\newblock \bibinfo{journal}{\emph{Neural computation}} \bibinfo{volume}{9},
  \bibinfo{number}{8} (\bibinfo{year}{1997}), \bibinfo{pages}{1735--1780}.
\newblock


\bibitem[\protect\citeauthoryear{Huang, Liu, Van Der~Maaten, and
  Weinberger}{Huang et~al\mbox{.}}{2017}]%
        {huang2017densely}
\bibfield{author}{\bibinfo{person}{Gao Huang}, \bibinfo{person}{Zhuang Liu},
  \bibinfo{person}{Laurens Van Der~Maaten}, {and} \bibinfo{person}{Kilian~Q
  Weinberger}.} \bibinfo{year}{2017}\natexlab{}.
\newblock \showarticletitle{Densely connected convolutional networks}. In
  \bibinfo{booktitle}{\emph{Proceedings of the IEEE conference on computer
  vision and pattern recognition}}. \bibinfo{pages}{4700--4708}.
\newblock


\bibitem[\protect\citeauthoryear{Jenlink}{Jenlink}{2006}]%
        {JenlinkKarenEmbry2006ME}
\bibfield{author}{\bibinfo{person}{Karen~Embry Jenlink}.}
  \bibinfo{year}{2006}\natexlab{}.
\newblock \bibinfo{title}{Math Education}.
\newblock , \bibinfo{numpages}{647--651}~pages.
\newblock
\showISBNx{978-1-4129-3958-4}


\bibitem[\protect\citeauthoryear{Junczys-Dowmunt and
  Grundkiewicz}{Junczys-Dowmunt and Grundkiewicz}{2016}]%
        {junczys2016phrase}
\bibfield{author}{\bibinfo{person}{Marcin Junczys-Dowmunt} {and}
  \bibinfo{person}{Roman Grundkiewicz}.} \bibinfo{year}{2016}\natexlab{}.
\newblock \showarticletitle{Phrase-based machine translation is
  state-of-the-art for automatic grammatical error correction}.
\newblock \bibinfo{journal}{\emph{arXiv preprint arXiv:1605.06353}}
  (\bibinfo{year}{2016}).
\newblock


\bibitem[\protect\citeauthoryear{Lave and Wenger}{Lave and Wenger}{1999}]%
        {lave1999legitimate}
\bibfield{author}{\bibinfo{person}{Jean Lave} {and} \bibinfo{person}{Etienne
  Wenger}.} \bibinfo{year}{1999}\natexlab{}.
\newblock \showarticletitle{Legitimate peripheral participation}.
\newblock \bibinfo{journal}{\emph{Learners, learning and assessment, London:
  The Open University}} (\bibinfo{year}{1999}), \bibinfo{pages}{83--89}.
\newblock


\bibitem[\protect\citeauthoryear{Levenshtein}{Levenshtein}{1966}]%
        {levenshtein1966binary}
\bibfield{author}{\bibinfo{person}{Vladimir~I Levenshtein}.}
  \bibinfo{year}{1966}\natexlab{}.
\newblock \showarticletitle{Binary codes capable of correcting deletions,
  insertions, and reversals}. In \bibinfo{booktitle}{\emph{Soviet physics
  doklady}}, Vol.~\bibinfo{volume}{10}. \bibinfo{pages}{707--710}.
\newblock


\bibitem[\protect\citeauthoryear{Li, Lu, Ding, and Gu}{Li
  et~al\mbox{.}}{2016}]%
        {Li2016PredictingCE}
\bibfield{author}{\bibinfo{person}{Guo Li}, \bibinfo{person}{Tun Lu},
  \bibinfo{person}{Xianghua Ding}, {and} \bibinfo{person}{Ning Gu}.}
  \bibinfo{year}{2016}\natexlab{}.
\newblock \showarticletitle{Predicting Collaborative Edits of Questions and
  Answers in Online Q\&A Sites}.
\newblock \bibinfo{journal}{\emph{Journal of Internet Technology}}
  \bibinfo{volume}{17} (\bibinfo{year}{2016}), \bibinfo{pages}{1187--1194}.
\newblock


\bibitem[\protect\citeauthoryear{Li, Zhu, Lu, Ding, and Gu}{Li
  et~al\mbox{.}}{2015}]%
        {li2015good}
\bibfield{author}{\bibinfo{person}{Guo Li}, \bibinfo{person}{Haiyi Zhu},
  \bibinfo{person}{Tun Lu}, \bibinfo{person}{Xianghua Ding}, {and}
  \bibinfo{person}{Ning Gu}.} \bibinfo{year}{2015}\natexlab{}.
\newblock \showarticletitle{Is it good to be like Wikipedia? Exploring the
  trade-offs of introducing collaborative editing model to Q\&A sites}. In
  \bibinfo{booktitle}{\emph{Proceedings of the 18th ACM Conference on Computer
  Supported Cooperative Work \& Social Computing}}.
  \bibinfo{pages}{1080--1091}.
\newblock


\bibitem[\protect\citeauthoryear{Likert}{Likert}{1932}]%
        {likert1932technique}
\bibfield{author}{\bibinfo{person}{Rensis Likert}.}
  \bibinfo{year}{1932}\natexlab{}.
\newblock \showarticletitle{A technique for the measurement of attitudes.}
\newblock \bibinfo{journal}{\emph{Archives of psychology}}
  (\bibinfo{year}{1932}).
\newblock


\bibitem[\protect\citeauthoryear{Ma, Xing, Chen, Chen, Qu, and Li}{Ma
  et~al\mbox{.}}{2019}]%
        {ma2019easy}
\bibfield{author}{\bibinfo{person}{Suyu Ma}, \bibinfo{person}{Zhenchang Xing},
  \bibinfo{person}{Chunyang Chen}, \bibinfo{person}{Cheng Chen},
  \bibinfo{person}{Lizhen Qu}, {and} \bibinfo{person}{Guoqiang Li}.}
  \bibinfo{year}{2019}\natexlab{}.
\newblock \showarticletitle{Easy-to-deploy API extraction by multi-level
  feature embedding and transfer learning}.
\newblock \bibinfo{journal}{\emph{IEEE Transactions on Software Engineering}}
  (\bibinfo{year}{2019}).
\newblock


\bibitem[\protect\citeauthoryear{Mamykina, Manoim, Mittal, Hripcsak, and
  Hartmann}{Mamykina et~al\mbox{.}}{2011}]%
        {MamykinaLena2011Dlft}
\bibfield{author}{\bibinfo{person}{Lena Mamykina}, \bibinfo{person}{Bella
  Manoim}, \bibinfo{person}{Manas Mittal}, \bibinfo{person}{George Hripcsak},
  {and} \bibinfo{person}{Björn Hartmann}.} \bibinfo{year}{2011}\natexlab{}.
\newblock \showarticletitle{Design lessons from the fastest q\&a site in the
  west}. In \bibinfo{booktitle}{\emph{Proceedings of the SIGCHI Conference on
  human factors in computing systems}} \emph{(\bibinfo{series}{CHI '11})}.
  \bibinfo{publisher}{ACM}, \bibinfo{pages}{2857--2866}.
\newblock
\showISBNx{9781450302289}


\bibitem[\protect\citeauthoryear{Middendorf}{Middendorf}{2018}]%
        {MiddendorfJessica2018IRtM}
\bibfield{author}{\bibinfo{person}{Jessica Middendorf}.}
  \bibinfo{year}{2018}\natexlab{}.
\newblock \bibinfo{title}{Increasing Retention through Math Study Skills}.
\newblock
\newblock
\showISBNx{978-0-438-08052-2}
\urldef\tempurl%
\url{http://search.proquest.com/docview/2061549897/}
\showURL{%
\tempurl}


\bibitem[\protect\citeauthoryear{Mikolov, Sutskever, Chen, Corrado, and
  Dean}{Mikolov et~al\mbox{.}}{2013}]%
        {mikolov2013distributed}
\bibfield{author}{\bibinfo{person}{Tomas Mikolov}, \bibinfo{person}{Ilya
  Sutskever}, \bibinfo{person}{Kai Chen}, \bibinfo{person}{Greg~S Corrado},
  {and} \bibinfo{person}{Jeff Dean}.} \bibinfo{year}{2013}\natexlab{}.
\newblock \showarticletitle{Distributed representations of words and phrases
  and their compositionality}. In \bibinfo{booktitle}{\emph{Advances in neural
  information processing systems}}. \bibinfo{pages}{3111--3119}.
\newblock


\bibitem[\protect\citeauthoryear{Mizumoto and Matsumoto}{Mizumoto and
  Matsumoto}{2016}]%
        {mizumoto2016discriminative}
\bibfield{author}{\bibinfo{person}{Tomoya Mizumoto} {and} \bibinfo{person}{Yuji
  Matsumoto}.} \bibinfo{year}{2016}\natexlab{}.
\newblock \showarticletitle{Discriminative reranking for grammatical error
  correction with statistical machine translation}. In
  \bibinfo{booktitle}{\emph{Proceedings of the 2016 Conference of the North
  American Chapter of the Association for Computational Linguistics: Human
  Language Technologies}}. \bibinfo{pages}{1133--1138}.
\newblock


\bibitem[\protect\citeauthoryear{Montoya, Ma, and Mondrag{\'o}n}{Montoya
  et~al\mbox{.}}{2013}]%
        {montoya2013social}
\bibfield{author}{\bibinfo{person}{Leydi~Viviana Montoya},
  \bibinfo{person}{Athen Ma}, {and} \bibinfo{person}{Ra{\'u}l~J
  Mondrag{\'o}n}.} \bibinfo{year}{2013}\natexlab{}.
\newblock \showarticletitle{Social achievement and centrality in MathOverflow}.
\newblock In \bibinfo{booktitle}{\emph{Complex Networks IV}}.
  \bibinfo{publisher}{Springer}, \bibinfo{pages}{27--38}.
\newblock


\bibitem[\protect\citeauthoryear{Napoles, Sakaguchi, Post, and
  Tetreault}{Napoles et~al\mbox{.}}{2016}]%
        {napoles2016gleu}
\bibfield{author}{\bibinfo{person}{Courtney Napoles}, \bibinfo{person}{Keisuke
  Sakaguchi}, \bibinfo{person}{Matt Post}, {and} \bibinfo{person}{Joel
  Tetreault}.} \bibinfo{year}{2016}\natexlab{}.
\newblock \showarticletitle{GLEU without tuning}.
\newblock \bibinfo{journal}{\emph{arXiv preprint arXiv:1605.02592}}
  (\bibinfo{year}{2016}).
\newblock


\bibitem[\protect\citeauthoryear{Ortiz-Mart{\i}nez, Garc{\i}a-Varea, and
  Casacuberta}{Ortiz-Mart{\i}nez et~al\mbox{.}}{2005}]%
        {ortiz2005thot}
\bibfield{author}{\bibinfo{person}{Daniel Ortiz-Mart{\i}nez},
  \bibinfo{person}{Ismael Garc{\i}a-Varea}, {and} \bibinfo{person}{Francisco
  Casacuberta}.} \bibinfo{year}{2005}\natexlab{}.
\newblock \showarticletitle{Thot: a toolkit to train phrase-based statistical
  translation models}.
\newblock \bibinfo{journal}{\emph{Tenth Machine Translation Summit. AAMT,
  Phuket, Thailand, September}} (\bibinfo{year}{2005}).
\newblock


\bibitem[\protect\citeauthoryear{Papineni, Roukos, Ward, and Zhu}{Papineni
  et~al\mbox{.}}{2002}]%
        {papineni-etal-2002-bleu}
\bibfield{author}{\bibinfo{person}{Kishore Papineni}, \bibinfo{person}{Salim
  Roukos}, \bibinfo{person}{Todd Ward}, {and} \bibinfo{person}{Wei-Jing Zhu}.}
  \bibinfo{year}{2002}\natexlab{}.
\newblock \showarticletitle{{B}leu: a Method for Automatic Evaluation of
  Machine Translation}. In \bibinfo{booktitle}{\emph{Proceedings of the 40th
  Annual Meeting of the Association for Computational Linguistics}}.
  \bibinfo{publisher}{Association for Computational Linguistics},
  \bibinfo{address}{Philadelphia, Pennsylvania, USA},
  \bibinfo{pages}{311--318}.
\newblock
\urldef\tempurl%
\url{https://www.aclweb.org/anthology/P02-1040}
\showURL{%
\tempurl}


\bibitem[\protect\citeauthoryear{Suzuki, Tamari, Fukuda, Uchida, and
  Kanahori}{Suzuki et~al\mbox{.}}{2003}]%
        {SuzukiMasakazu2003IaiO-infty}
\bibfield{author}{\bibinfo{person}{Masakazu Suzuki}, \bibinfo{person}{Fumikazu
  Tamari}, \bibinfo{person}{Ryoji Fukuda}, \bibinfo{person}{Seiichi Uchida},
  {and} \bibinfo{person}{Toshihiro Kanahori}.} \bibinfo{year}{2003}\natexlab{}.
\newblock \showarticletitle{INFTY: an integrated OCR system for mathematical
  documents}. In \bibinfo{booktitle}{\emph{Proceedings of the 2003 ACM
  symposium on document engineering}} \emph{(\bibinfo{series}{DocEng '03})}.
  \bibinfo{publisher}{ACM}, \bibinfo{pages}{95--104}.
\newblock
\showISBNx{1581137249}


\bibitem[\protect\citeauthoryear{Tausczik, Kittur, and Kraut}{Tausczik
  et~al\mbox{.}}{2014}]%
        {tausczik2014collaborative}
\bibfield{author}{\bibinfo{person}{Yla~R Tausczik}, \bibinfo{person}{Aniket
  Kittur}, {and} \bibinfo{person}{Robert~E Kraut}.}
  \bibinfo{year}{2014}\natexlab{}.
\newblock \showarticletitle{Collaborative problem solving: A study of
  mathoverflow}. In \bibinfo{booktitle}{\emph{Proceedings of the 17th ACM
  conference on Computer supported cooperative work \& social computing}}.
  \bibinfo{pages}{355--367}.
\newblock


\bibitem[\protect\citeauthoryear{Twaakyondo and Okamoto}{Twaakyondo and
  Okamoto}{1995}]%
        {twaakyondo1995structure}
\bibfield{author}{\bibinfo{person}{Hashim~M Twaakyondo} {and}
  \bibinfo{person}{Masayuki Okamoto}.} \bibinfo{year}{1995}\natexlab{}.
\newblock \showarticletitle{Structure analysis and recognition of mathematical
  expressions}. In \bibinfo{booktitle}{\emph{Proceedings of 3rd International
  Conference on Document Analysis and Recognition}}, Vol.~\bibinfo{volume}{1}.
  IEEE, \bibinfo{pages}{430--437}.
\newblock


\bibitem[\protect\citeauthoryear{Vargo and Matsubara}{Vargo and
  Matsubara}{2016}]%
        {vargo2016editing}
\bibfield{author}{\bibinfo{person}{Andrew~W Vargo} {and}
  \bibinfo{person}{Shigeo Matsubara}.} \bibinfo{year}{2016}\natexlab{}.
\newblock \showarticletitle{Editing Unfit Questions in Q\&A}. In
  \bibinfo{booktitle}{\emph{2016 5th IIAI International Congress on Advanced
  Applied Informatics (IIAI-AAI)}}. IEEE, \bibinfo{pages}{107--112}.
\newblock


\bibitem[\protect\citeauthoryear{Vaswani, Shazeer, Parmar, Uszkoreit, Jones,
  Gomez, Kaiser, and Polosukhin}{Vaswani et~al\mbox{.}}{2017}]%
        {vaswani2017attention}
\bibfield{author}{\bibinfo{person}{Ashish Vaswani}, \bibinfo{person}{Noam
  Shazeer}, \bibinfo{person}{Niki Parmar}, \bibinfo{person}{Jakob Uszkoreit},
  \bibinfo{person}{Llion Jones}, \bibinfo{person}{Aidan~N Gomez},
  \bibinfo{person}{{\L}ukasz Kaiser}, {and} \bibinfo{person}{Illia
  Polosukhin}.} \bibinfo{year}{2017}\natexlab{}.
\newblock \showarticletitle{Attention is all you need}. In
  \bibinfo{booktitle}{\emph{Advances in neural information processing
  systems}}. \bibinfo{pages}{5998--6008}.
\newblock


\bibitem[\protect\citeauthoryear{Wang, Sun, and Wang}{Wang
  et~al\mbox{.}}{2019b}]%
        {wang2019image}
\bibfield{author}{\bibinfo{person}{Jian Wang}, \bibinfo{person}{Yunchuan Sun},
  {and} \bibinfo{person}{Shenling Wang}.} \bibinfo{year}{2019}\natexlab{b}.
\newblock \showarticletitle{Image To Latex with DenseNet Encoder and Joint
  Attention}.
\newblock \bibinfo{journal}{\emph{Procedia computer science}}
  \bibinfo{volume}{147} (\bibinfo{year}{2019}), \bibinfo{pages}{374--380}.
\newblock


\bibitem[\protect\citeauthoryear{Wang, Chen, and Xing}{Wang
  et~al\mbox{.}}{2019a}]%
        {wang2019domain}
\bibfield{author}{\bibinfo{person}{Xu Wang}, \bibinfo{person}{Chunyang Chen},
  {and} \bibinfo{person}{Zhenchang Xing}.} \bibinfo{year}{2019}\natexlab{a}.
\newblock \showarticletitle{Domain-specific machine translation with recurrent
  neural network for software localization}.
\newblock \bibinfo{journal}{\emph{Empirical Software Engineering}}
  \bibinfo{volume}{24}, \bibinfo{number}{6} (\bibinfo{year}{2019}),
  \bibinfo{pages}{3514--3545}.
\newblock


\bibitem[\protect\citeauthoryear{Xu, Ba, Kiros, Cho, Courville, Salakhudinov,
  Zemel, and Bengio}{Xu et~al\mbox{.}}{2015}]%
        {xu2015show}
\bibfield{author}{\bibinfo{person}{Kelvin Xu}, \bibinfo{person}{Jimmy Ba},
  \bibinfo{person}{Ryan Kiros}, \bibinfo{person}{Kyunghyun Cho},
  \bibinfo{person}{Aaron Courville}, \bibinfo{person}{Ruslan Salakhudinov},
  \bibinfo{person}{Rich Zemel}, {and} \bibinfo{person}{Yoshua Bengio}.}
  \bibinfo{year}{2015}\natexlab{}.
\newblock \showarticletitle{Show, attend and tell: Neural image caption
  generation with visual attention}. In \bibinfo{booktitle}{\emph{International
  conference on machine learning}}. \bibinfo{pages}{2048--2057}.
\newblock


\bibitem[\protect\citeauthoryear{Yuan and Briscoe}{Yuan and Briscoe}{2016}]%
        {yuan2016grammatical}
\bibfield{author}{\bibinfo{person}{Zheng Yuan} {and} \bibinfo{person}{Ted
  Briscoe}.} \bibinfo{year}{2016}\natexlab{}.
\newblock \showarticletitle{Grammatical error correction using neural machine
  translation}. In \bibinfo{booktitle}{\emph{Proceedings of the 2016 Conference
  of the North American Chapter of the Association for Computational
  Linguistics: Human Language Technologies}}. \bibinfo{pages}{380--386}.
\newblock


\bibitem[\protect\citeauthoryear{Yuan, Briscoe, and Felice}{Yuan
  et~al\mbox{.}}{2016}]%
        {yuan2016candidate}
\bibfield{author}{\bibinfo{person}{Zheng Yuan}, \bibinfo{person}{Ted Briscoe},
  {and} \bibinfo{person}{Mariano Felice}.} \bibinfo{year}{2016}\natexlab{}.
\newblock \showarticletitle{Candidate re-ranking for SMT-based grammatical
  error correction}. In \bibinfo{booktitle}{\emph{Proceedings of the 11th
  Workshop on Innovative Use of NLP for Building Educational Applications}}.
  \bibinfo{pages}{256--266}.
\newblock


\end{thebibliography}
\received{October 2020 }
\received[revised]{April 2021 }
\received[accepted]{July 2021} 

\end{document}